\documentclass[fleqn,usenatbib]{mnras}

\usepackage{newtxtext,newtxmath}
\usepackage[T1]{fontenc}
\usepackage{ae,aecompl}
\usepackage{adjustbox}
\usepackage{placeins}
\usepackage{enumitem}
\usepackage{caption}
\usepackage{comment}
\usepackage{float}
\usepackage{subcaption}
\usepackage{hyperref}
\newcommand{\RNum}[1]{\uppercase\expandafter{\romannumeral #1\relax}}
\usepackage{pdfpages}

\usepackage{graphicx}	% Including figure files
\usepackage{amsmath}	% Advanced maths commands
\usepackage{amssymb}	% Extra maths symbols

\title[Timing of young radio pulsars I] {Timing of young radio pulsars $\rm{\RNum{1}}$: Timing noise,   periodic modulation and proper motion.}

\author[A. Parthasarathy et al.]
{A. Parthasarathy$^{1,2,3}$,\thanks{E-mail: aparthas@swin.edu.au}
R.M. Shannon$^{1,2}$,
S. Johnston$^{3}$,
L. Lentati$^{4}$,
M. Bailes$^{1,2}$, \newauthor 
S. Dai$^{3}$, 
M. Kerr$^{5}$, 
R.N. Manchester$^{3}$,
S. Oslowski$^{1,2}$,
C. Sobey$^{6}$, \newauthor
W. van Straten$^{7}$,
P. Weltevrede$^{8}$
\\
$^{1}$ Centre for Astrophysics and Supercomputing, Swinburne University of Technology, P.O. Box 218, Hawthorn, Victoria 3122, Australia \\
$^{2}$ OzGrav: Australian Research Council Centre of Excellence for Gravitational Wave Discovery \\
$^{3}$ CSIRO Astronomy and Space Science, Australia Telescope National Facility, PO~Box~76, Epping NSW~1710, Australia \\
$^{4}$ Astrophysics Group, Cavendish Laboratory, JJ Thomson Avenue,  Cambridge, CB3 0HE, UK \\
$^{5}$ Space Science Division, Naval Research Laboratory, Washington, DC 20375, USA \\
$^{6}$ CSIRO Astronomy and Space Science, PO Box 1130 Bentley, WA 6102, Australia \\
$^{7}$ Institute for Radio Astronomy \& Space Research, Auckland University of Technology, Private Bag 92006, Auckland 1142, New Zealand \\
$^{8}$ Jodrell Bank Centre for Astrophysics, The University of Manchester, Alan Turing Building, Manchester, M13 9PL, United Kingdom \\ 
}

\date{Accepted XXX. Received YYY; in original form ZZZ}

\pubyear{2019}

\begin{document}
\label{firstpage}
\pagerange{\pageref{firstpage}--\pageref{lastpage}}
\maketitle

\begin{abstract}
The smooth spin-down of young pulsars is perturbed by two non-deterministic phenomenon, glitches and timing noise. Although the timing noise provides insights into nuclear and plasma physics at extreme densities, it acts as a barrier to high-precision pulsar timing experiments. %often leads to biased measurements of long-term spin-down and astrometric parameters. 
An improved methodology based on Bayesian inference is developed to simultaneously model the stochastic and deterministic parameters for a sample of 85 high-$\dot{E}$ radio pulsars observed for $\sim$ 10 years with the 64-m Parkes radio telescope. Timing noise is known to be a red process and we develop a parametrization based on the red-noise amplitude ($A_{\rm red}$) and spectral index ($\beta$). We measure the median $A_{\rm red}$ to be $-10.4^{+1.8}_{-1.7}$ yr$^{3/2}$ and $\beta$ to be $-5.2^{+3.0}_{-3.8}$ and show that the strength of timing noise scales proportionally to $\nu^{1}|\dot{\nu}|^{-0.6\pm0.1}$, where $\nu$ is the spin frequency of the pulsar and $\dot{\nu}$ its spin-down rate. Finally, we measure significant braking indices for 19 pulsars, proper motions for two pulsars and discuss the presence of periodic modulation in the arrival times of five pulsars.
\end{abstract}


%%%%%%%%%%%%%%%%%%%%%%%%%%%%%%%%%%%%%%%%%%%%%%%%%%

%%%%%%%%%%%%%%%%% BODY OF PAPER %%%%%%%%%%%%%%%%%%

\section{Introduction}
Young neutron stars provide unique insights into astrophysics that are not available from the bulk of the pulsar population. They frequently exhibit two types of deviations from a steady spin-down behaviour, `glitches' and `timing noise'. Glitches are sudden jumps in the pulsars' spin-frequency acting as probes of neutron star interiors. Timing noise is a type of rotational irregularity which causes the pulse arrival times to stochastically wander about a steady spin-down state. Our sample is representative of pulsars that are spinning down rapidly and present the most promising avenue for detailed studies of timing noise, glitches and their spin-down behaviour.  

The technique of pulsar timing enables the precise measurement of their spin periods ($P$) and their spin-down rates ($\dot{P}$), allowing us to study their evolution in the $P-\dot{P}$ diagram (\citealt{ppdot_evolution}), see Figure 1. Although young pulsar timing offers several opportunities to explore a plethora of astrophysical phenomena, it is a challenging prospect as most of these astrophysical signals are dominated or biased by timing noise and glitches. A careful methodology is thus needed in the analysis of young pulsar timing data to disentangle the deterministic processes from the stochastic components. For example, young pulsars are thought to be associated with supernova remnants, and measuring their proper motions (\citealt{hobbs_propermotion}) allows us to probe the connections between the neutron star and its progenitor, which has implications for birthrate statistics (\citealt{birthratestats}). Unbiased measurements of proper motion through pulsar timing can be obtained only if the timing noise in the pulse arrival times is modelled accurately. While understanding the origin of the stochastic signals present in the ToAs is important, it is also essential to characterize and mitigate the effects of these signals as part of the general timing model because it reduces the bias in the estimation of other deterministic pulsar parameters. 

  \begin{figure}
  \centering
  \includegraphics[angle=0,width=0.45\textwidth]{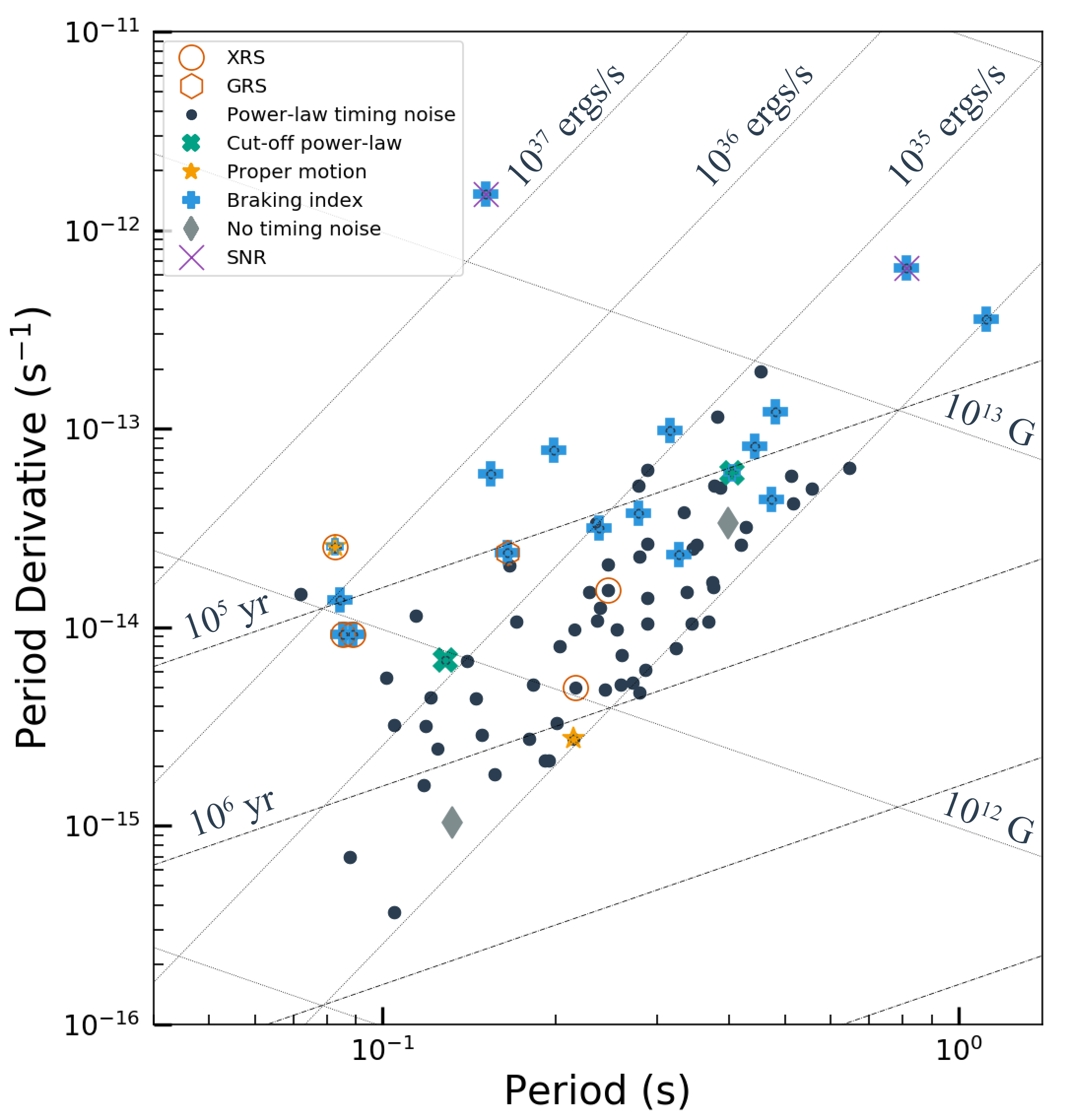}
  \caption{\label{fig:ppdot} $P-\dot{P}$ diagram showing our sample of 85 young pulsars coloured according to their preferred timing model. The different timing models are outlined in Section \ref{timing analysis}. A few pulsars are also highlighted to be X-ray (XRS) or Gamma-ray sources (GRS) and/or to have known supernova associations (SNR). Our sample of pulsars mostly have $\dot{E} > 10^{34}$ ergs/s, surface magnetic fields ranging from $10^{\rm 12}G$ to $10^{\rm 13}G$ with characteristic ages of $10^{\rm 5}$ to $10^{\rm 6}$ years.}
  \end{figure}

\subsection{Timing noise}
Timing noise manifests itself as a red-noise process in the ToAs, implying an autocorrelated process on a time-scale of months to years and is generally described by a wide-sense stationary stochastic signal (\citealt{groth_timingnoise}). \cite{boynton_timingnoise} attempted to describe the timing noise in the Crab pulsar as random walks in either the phase, frequency or the spin-down parameter of the pulsar. They showed that the power spectra expected from such random walks will be proportional to $-2$, $-4$ and $-6$ for phase, frequency and spin-down respectively. Following this, many attempts have been made to study the timing noise in pulsars over increasing data spans and for a larger sample of pulsars. \cite{cordeshelfand_timingnoise} studied the timing behaviour of 50 pulsars and found that the timing activity was correlated with $\dot{P}$ but weakly correlated with $P$ and concluded that timing activity is consistent with a random walk origin. As more pulsars with longer data sets were studied, it became apparent that timing noise might be explained by a combination of different random walks in pulsar spin-frequency ($\nu$) and spin-frequency derivative ($\dot{\nu}$) and by discrete jumps in phase and spin-parameters. Timing noise is thought to arise due to changes in the coupling between the neutron star crust and its super-fluid core (\citealt{superfluid_timingnoise}) or magnetospheric torque fluctuations (\citealt{glitches_timingnoise}, \citealt{magnetospheric_timingnoise}). It has also been attributed to microjumps, which are similar to small glitches (\citealt{melatos_microglitches}) and fluctuations in the spin-down torque (\citealt{cheng_timingnoise}).  It has often been suggested that the superfluid interior of a neutron star can have macroscopic Kolmogorov-like turbulence which can contribute to stochasticity in the spin-down processes observed in radio pulsars (\citealt{greenstein_turbulence}, \citealt{link_turbulence}, \citealt{melatos_link}). 

The observations of quasi-periodic state switching of pulsars (\citealt{kramermagnet_timingnoise}, \citealt{magnetospheric_timingnoise}), each state with a distinct spin-down rate, led to alternative descriptions of timing noise being periodic or quasi-periodic processes. Unmodelled planetary companions (\citealt{mkerr_planets}), pulse-shape changes (\citealt{brook_profilechanges}), accretion from the ISM (\citealt{cordestheory_timingnoise}) or free precession (\citealt{lyne_precession}, \citealt{mkerr_precession}) have also been attributed as explanations for the observed low-frequency structures in the ToAs. \cite{hobbs_propermotion}, studied a large sample of pulsars observed over $\sim$ 10 years and concluded that timing noise is widespread in pulsars, and that it cannot be explained as a simple random walk in pulse phase, frequency or spin-down rate. The timing noise in millisecond pulsars (MSPs) has been mainly studied to understand their sensitivity to nHz-frequency gravitational waves (\citealt{lam_msp}, \citealt{callabero_msp}, \citealt{lentati_2016}). However, unlike millisecond pulsars, the timing noise in young pulsars is very strong, often contributing many cycles of pulse phase on week to month timescales.  \cite{shannon_timingnoise} pointed out that the observed strength of timing noise varies by more than eight orders of magnitude over magnetars, young and millisecond pulsars.

\subsection{Pulsar spin-down and braking index}
The long-term spin-down of a pulsar can be approximated as
  \begin{equation} \label{spin_down}
  \dot{\nu} = -K\nu^{\rm n},
  \end{equation}
where K is a constant and $n$ is the braking index. The braking index describes the relationship between the braking torque acting on a pulsar and its spin-frequency parameters, and provides a probe into the physics dictating pulsar temporal evolution.
We solve for $n$ by taking the derivative of equation \ref{spin_down},
  \begin{equation} \label{brakingindex}
  n = \frac{\nu\ddot{\nu}}{\dot{\nu}^2},
  \end{equation}
where $\ddot{\nu}$ is the second derivative of the spin frequency. For standard magnetic-dipole braking, the magnetic field strength and the magnetic-dipole inclination angle are assumed to be constant in time, with $n=3$ (\citealt{glitch_template}).  While measuring $\nu$ and $\dot{\nu}$ is trivial using standard timing methods, measuring the long-term $\ddot{\nu}$ is challenging, mainly because of the fact that it is a very small quantity. In `old' pulsars, with $\nu \sim$ 1 Hz and $\dot{\nu} \sim$ $10^{-15} \,\rm{Hz}\,s^{-1}$, the estimated $\ddot{\nu}$ from equation \ref{brakingindex} is $\sim 10^{-30}\,\rm{Hz}\,s^{-2}$. However for the youngest pulsars we estimate $\ddot{\nu}$ to be $< 10^{-20}\,\rm{Hz}\,s^{-2}$, which makes these pulsars suitable for studying pulsar braking mechanisms (\citealt{johnstongalloway_braking1}). If both $K$ and $n$ are constant in time, a pulsar will follow a track in the $P-\dot{P}$ diagram with a slope of $2-n$. The $P-\dot{P}$ diagram can then be used as an evolutionary tool in which pulsars are born in the upper-left region and as they age and spin-down, they drift towards the cluster of ``normal'' pulsars, with periods of $\sim$ 0.5 s (\citealt{ppdot_evolution}). 

Both timing noise and glitches introduce variations in $\dot{\nu}$ which becomes problematical in the long-term measurement of $\ddot{\nu}$. Glitches are often modelled as permanent changes in spin-frequency ($\nu$) and spin-frequency derivative ($\dot{\nu}$) or as exponential decays in $\nu$ over $\tau$ days and are typically attributed to either the transfer of angular momentum between the super-fluid interior and the solid crust of the neutron star (\citealt{glitchtheory}, \citealt{glitchtheory2}) or as star quakes in the crystalline outer crust of the neutron star (\citealt{glitchtheory3}).  

\subsection{Quasi-Periodic modulations}
The reflex motion resulting from the orbital motion of a companion to a pulsar, introduces modulations in the ToAs, which led to e.g. the discovery of the double neutron star system B1913+16 (\citealt{hulsetaylor}) and the first exoplanets (\citealt{firstexoplanet}). 
%Other effects that can lead to ToA modulation include the precession of the neutron star and reconfiguration of the pulsar magnetosphere. 
Precession induces a periodic change in the spin-down torque which causes ToA modulation and since our line of sight cuts across different parts of the neutron star polar cap, there can also be an observed change in the shape of the pulse profile (\citealt{precession_interpretation}). Such events of ToA modulations were reported by \cite{magnetospheric_timingnoise} in 17 pulsars, of which 6 showed correlations with pulse profile variations. Recently, \cite{stairs_2019} reported correlated shape and spin-down changes in PSR J1830--1059, which they attributed to large-scale magnetospheric switching.
%However, since the observed pulse profile seemed to oscillate between two distinct states, precession of the neutron star did not seem plausible. Instead, this was
\cite{brook_profilechanges} analysed 168 pulsars and searched for correlations between profile shape changes and $\dot{\nu}$ and found that although this correlation is clear in some pulsars, the intrinsic relationship between change in $\dot{\nu}$ and profile variability may be much more complex than previously postulated (see also \citealt{mkerr_precession}).

\subsection{Proper motions}
Pulsars are created in supernovae, and the birth process is expected to impart a high `kick velocity'. Various mechanisms have been proposed for these kicks, including an asymmetric neutrino emission in the presence of super-strong magnetic fields \citep{SNTheory_neutrino}, a postnatal electromagnetic rocket mechanism \citep{SNTheory_postnatal}, asymmetric explosion of $\gamma$-ray bursts \citep{SNTheory_gamma} and hydrodynamical instabilities in the collapsed supernova core \citep{SNTheory_hydrodynamics}. While the evidence for such kicks is unequivocal (\citealt{johnston_propermotion}), their physical origin remains unclear. A pulsar's proper motion causes sinusoidal variations in ToAs with a periodicity of one year and an amplitude which increases with time.
%\cite{gunn_propermotions} concluded that the population of young neutron stars receive a velocity averaging 100 km/s at birth, making these velocities significantly larger than their progenitors high-mass OB stars (\citealt{hobbs_propermotion}). Measuring pulsar proper motions and velocities allows us to constrain some of the proposed theories and also helps determine the birth rate of pulsars and their associations with supernova remnants. \cite{johnston_propermotion} measured the velocity vectors for 25 pulsars and concluded that there is strong evidence for the alignment of the rotation and velocity vectors. 
%As data sets increased, it became clear that the presence of timing noise could bias the measured proper motion values.

\subsection{The Bayesian pulsar timing framework}
\cite{lentati_2013} pointed out that in order to obtain an unbiased estimation of the pulsar parameters (proper motions, spin parameters, braking index etc.) it is important to simultaneously model the stochastic (timing noise) and the deterministic (pulsar) parameters. Most of the frequentist approaches (\citealt{hobbs_whitening}, \citealt{coles_whitening}) do not consider the covariances between the timing model and the stochastic processes, and the uncertainties in the parameter estimates, motivating the development of \textsc{temponest} (\citealt{temponest}), which performs a simultaneous analysis of the timing model and additional stochastic parameters using the Bayesian inference tool, \textsc{Multinest} (\citealt{multinest}, \citealt{multnest2}). It also allows for robust model selection between different sets of timing parameters based on the Bayesian evidences.
We use \textsc{temponest} to simultaneously model the pulsar parameters and the noise parameters 
%for 85 high $\dot{E}$, young pulsars observed with the CSIRO Parkes radio telescope over a span of $\sim$ 10 years 
and use the Bayesian evidence to select the optimal model for each pulsar. Such an analysis allows us to discuss the statistical properties of timing noise and also compare the results with those obtained from other Bayesian tools. 

In Section \ref{Observations}, we describe the observing program and the data processing pipeline. In Section \ref{timing analysis}, we describe the Bayesian timing analysis in detail and present the mathematical formulation of the timing model. In Section \ref{results}, we present the basic observational characteristics, the timing solutions, the timing noise models for our sample of pulsars along with the new proper motions. Finally in Section \ref{discussion}, we delve into the implications of our results.

\section{Observations} \label{Observations}
In this paper, we study 85 pulsars observed at a monthly cadence using the 64-m CSIRO Parkes radio telescope in support of the \textit{Fermi} mission that commenced in February 2007 (\citealt{pulsarsample}, \citealt{pulsarsample1}).  We selected pulsars for which there were no identified glitches \footnote{In subsequent analysis described below, two relatively small glitches were detected and parametrized.}. These pulsars have $\dot{E} > 10^{34}$ ergs/s, surface magnetic fields typically ranging from $10^{12}$G to $10^{13}$G with characteristic ages of $10^{5}$ to $10^{6}$ years as shown in Figure \ref{fig:ppdot}. Two pulsars, PSR ~J1513--5908 and J1632--4818 have known associations with supernova remnants, five other pulsars, PSR ~J0543+2329, J1224--6407, J1509--5850, J1809--1917, J1833--0827 are known X-ray sources and 3 others, J1509--5850, J1513--5908, J1648--4611 are known gamma-ray sources (\citealt{secondfermicat}). 

Most of these observations were carried out using the 20-cm multi-beam receiver (\citealt{mb_20cm}), with 256 MHz of bandwidth divided into 1024 frequency channels and folded in real-time into 1024 phase bins. Some of these pulsars were also observed at radio wavelengths of 10-cm and 40-cm. Each pulsar is observed for a few minutes depending upon its flux density. The observations were excised of radio frequency interference (RFI) and calibrated using standard \textsc{PSRCHIVE} (\citealt{psrchive}) tools and averaged in frequency, time and polarization. The ToAs of the pulses were computed by correlating a high signal-to-noise ratio, smoothed template with the averaged observations. For this analysis, we use only the 20-cm observations as the 10-cm data are sparsely spaced in time and the 50-cm data are highly corrupted by RFI. 

\begin{table*}
\caption{\label{obs_char_table}
Observational characteristics of the 85 pulsars described in this paper. The position and spin-down parameters are reported at the mentioned period epoch (PEPOCH) along with the timespan and the MJD range. The 95\% confidence limits for the position and spin-parameters reported here are derived from the preferred model for each pulsar. The confidence regions are individually stated, if the upper and lower confidence limits are asymmetric. }
\centering
\begin{adjustbox}{angle=0}
\renewcommand{\arraystretch}{1.5}
\resizebox{\textwidth}{!}{
\begin{tabular}{lrrrrrrrrrrr}
\hline
\hline
PSR & RAJ & DECJ & PEPOCH & $\nu$ & $\dot{\nu}$ & $N_{ToA}$ & Timespan & MJD Range   \\
 & (h:m:s) & (d:m:s) & & $(s^{-1})$ & $(10^{-14} s^{-2})$ & & (yr) & \\
\hline
J0543+2329 & {05:43:11.26}$^{0.05}_{0.62}$ & {23:16:39.66}$^{0.91}_{0.03}$ & 55580 & 4.06531029396(8) & -25.48351(13) & 111 & 9.6 & 54505-58011 \\
J0745--5353 & 07:45:04.48(4) & -53:53:09.56(3) & 55129 & 4.65465907222(10) & -4.73802(14) & 173 & 10.5 & 53973-57824 \\
J0820--3826 & 08:20:59.929(9) & -38:26:42.9(13) & 55583 & 8.01046656802(3) & -15.6734(5) & 115 & 9.0 & 54548-57824 \\
J0834--4159 & 08:34:17.807(2) & -41:59:35.99(2) & 55308 & 8.25642751376(12) & -29.18213(2) & 134 & 9.9 & 54220-57824 \\
J0857--4424 & 08:57:55.832(2) & -44:24:10.65(2) & 55335 & 3.0601045423(4) & -19.6145(10) & 170 & 9.9 & 54220-57824 \\
J0905--5127 & 09:05:51.96(2) & -51:27:54.05(2) & 55341 & 2.88766003664(2) & -20.7322(6) & 136 & 10.5 & 53971-57824 \\
J0954--5430 & 09:54:06.046(5) & -54:30:52.82(4) & 55323 & 2.11483307064(18) & -19.6358(5) & 125 & 9.9 & 54220-57824 \\
J1016--5819 & 10:16:12.071(2) & -58:19:01.07(16) & 55333 & 11.38507898552(7) & -9.05763(13) & 128 & 9.9 & 54220-57824 \\
J1020--6026 & 10:20:11.41(19) & -60:26:06.3(12) & 55494 & 7.11838566803(5) & -34.1421(5) & 81 & 6.4 & 54365-56708 \\
J1043--6116 & 10:43:55.261(2) & -61:16:50.76(2) & 55358 & 3.46493998447(4) & -12.49169(7) & 131 & 9.9 & 54220-57824 \\
J1115--6052 & 11:15:53.722(4) & -60:52:18.61(3) & 55366 & 3.84942036520(10) & -10.70996(13) & 130 & 10.4 & 54220-58011 \\
J1123--6259 & 11:23:55.53(12) & -62:59:10.94(8) & 55393 & 3.68410549479(18) & -7.13560(2) & 131 & 10.4 & 54220-58011 \\
J1156--5707 & 11:56:07.45(7) & -57:07:02.1(6) & 55354 & 3.4672047206(6) & -31.9149(9) & 134 & 10.4 & 54220-58011 \\
J1216--6223 & 12:16:41.96(13) & -62:23:57.00(9) & 55391 & 2.673417841032(14) & -12.02408(14) & 90 & 6.8 & 54220-56708 \\
J1224--6407 & 12:24:22.254(6) & -64:07:53.87(4) & 55191 & 4.61936868862(6) & -10.56960(8) & 274 & 10.4 & 54204-58011 \\
J1305--6203 & 13:05:21.14(10) & -62:03:21.07(8) & 55390 & 2.33768482203(6) & -17.57335(10) & 127 & 10.4 & 54220-58011 \\
J1349--6130 & 13:49:36.62(18) & -61:30:17.12(15) & 55429 & 3.85557384197(19) & -7.60678(3) & 171 & 10.4 & 54220-58012 \\
J1412--6145 & 14:12:07.63(10) & -61:45:28.48(8) & 55363 & 3.1720007909(13) & -99.643(4) & 162 & 10.4 & 54220-58012 \\
J1452--5851 & 14:52:52.60(10) & -58:51:13.2(11) & 55367 & 2.586365680859(2) & -33.91606(17) & 75 & 6.8 & 54220-56708 \\
J1453--6413 & 14:53:32.665(6) & -64:13:16.00(5) & 55433 & 5.57144021375(9) & -8.51812(15) & 184 & 10.4 & 54220-58012 \\
J1509--5850 & 15:09:27.156(7) & -58:50:56.01(8) & 55378 & 11.2454488757(7) & -115.9175(16) & 129 & 10.4 & 54220-58012 \\
J1512--5759 & 15:12:43.04(10) & -57:59:59.8(11) & 55383 & 7.77009392040(18) & -41.37272(2) & 131 & 10.4 & 54220-58012 \\
J1513--5908 & {15:13:55.81}$^{0.11}_{0.1}$ & {-59:08:09.64}$^{0.04}_{0.11}$ & 55336 & 6.59709182778(19) & -6653.10558(27) & 151 & 11.6 & 54220-58469 \\
J1514--5925 & 15:14:59.10(3) & -59:25:43.5(3) & 55415 & 6.72054447215(8) & -13.0014(17) & 85 & 6.8 & 54220-56708 \\
J1515--5720 & 15:15:09.23(14) & -57:20:50.15(17) & 55380 & 3.48859614104(17) & -7.41624(2) & 130 & 10.4 & 54220-58012 \\
J1524--5706 & 15:24:21.42(12) & -57:06:34.64(15) & 55383 & 0.89591729463(9) & -28.60366(2) & 128 & 10.4 & 54220-58012 \\
J1530--5327 & 15:30:26.892(2) & -53:27:56.02(4) & 55431 & 3.58476370133(5) & -6.01385(9) & 158 & 10.4 & 54220-58012 \\
J1531--5610 & 15:31:27.901(11) & -56:10:55.33(13) & 55304 & 11.8756292823(4) & -194.5360(14) & 140 & 10.4 & 54220-58012 \\
J1538--5551 & 15:38:45.016(5) & -55:51:36.95(8) & 55421 & 9.55329718930(4) & -29.2693(6) & 85 & 6.8 & 54220-56708 \\
J1539--5626 & 15:39:14.06(18) & -56:26:26.3(2) & 55408 & 4.10854528747(18) & -8.18323(2) & 128 & 10.4 & 54220-58012 \\
J1543--5459 & 15:43:56.43(6) & -54:59:15.0(8) & 55408 & 2.6515508603(4) & -36.6285(7) & 128 & 10.4 & 54220-58012 \\
J1548--5607 & 15:48:44.015(8) & -56:07:34.3(10) & 55408 & 5.85007580447(19) & -36.73172(3) & 128 & 10.4 & 54220-58012 \\
J1549--4848 & 15:49:21.08(17) & -48:48:35.5(3) & 55407 & 3.46794867500(2) & -16.96693(3) & 130 & 10.4 & 54220-58012 \\
J1551--5310 & 15:51:41.0(10) & -53:11:00.5(4) & 55383 & 2.20532573802(11) & -94.7569(18) & 84 & 6.8 & 54220-56708 \\
J1600--5751 & 16:00:19.90(11) & -57:51:15.3(13) & 55377 & 5.14255433375(2) & -5.63069(3) & 129 & 10.4 & 54220-58012 \\
J1601--5335 & 16:01:54.81(2) & -53:35:44.1(4) & 55391 & 3.46645281446(7) & -74.9184(10) & 86 & 6.8 & 54220-56708 \\
J1611--5209 & 16:11:03.37(01) & {-52:09:22.13}$^{0.1}_{0.11}$ & 55390 & 5.47960812333(12) & -15.52478(19) & 128 & 10.4 & 54220-58012 \\
J1632--4757 & 16:32:16.66(13) & -47:57:34.5(3) & 55419 & 4.37505274487(4) & -28.8454(7) & 83 & 6.8 & 54220-56708 \\
J1632--4818 & 16:32:39.70(3) & -48:18:53.8(8) & 55426 & 1.2289964712(14) & -98.0730(3) & 113 & 10.4 & 54220-58012 \\
J1637--4553 & 16:37:58.692(4) & -45:53:26.82(9) & 55443 & 8.41939738252(19) & -22.6194(4) & 159 & 11.1 & 53971-58012 \\
J1637--4642 & 16:37:13.75(17) & -46:42:14.2(4) & 55398 & 6.491542203(4) & -249.892(10) & 128 & 10.4 & 54220-58012 \\
\hline
\end{tabular}}
\end{adjustbox}
\end{table*}

\begin{table*}
\ContinuedFloat 
\caption{
Observational characteristics of the 85 pulsars described in this paper. The position and spin-down parameters are reported at the mentioned period epoch (PEPOCH) along with the timespan and the MJD range. The 95\% confidence limits for the position and spin-parameters reported here are derived from the preferred model for each pulsar.  The confidence regions are individually stated, if the upper and lower confidence limits are asymmetric (continued).}
\centering
\begin{adjustbox}{angle=0}
\renewcommand{\arraystretch}{1.5}
\resizebox{\textwidth}{!}{
\begin{tabular}{lrrrrrrrrrrr}
\hline
\hline
JName & RAJ & DECJ & PEPOCH & $\nu$ & $\dot{\nu}$ & $N_{ToA}$ & Timespan & MJD Range  \\
 & (h:m:s) & (d:m:s) & & $(s^{-1})$ & $(10^{-14} s^{-2})$ & & (yr) &  \\
\hline
J1638--4417 & 16:38:46.226(8) & -44:17:03.2(2) & 55410 & 8.4888197965(4) & -11.5716(7) & 128 & 10.4 & 54220-58012 \\
J1638--4608 & 16:38:23.26(9) & -46:08:13.4(3) & 55408 & 3.5951208300(10) & -66.5397(16) & 129 & 10.4 & 54220-58012 \\
J1640--4715 & 16:40:13.09(3) & -47:15:38.1(8) & 55392 & 1.9326284729(4) & -15.7266(6) & 128 & 10.4 & 54220-58012 \\
J1643--4505 & 16:43:36.91(3) & -45:05:45.8(7) & 55580 & 4.212470392(4) & -56.473(10) & 116 & 9.6 & 54505-58012 \\
J1648--4611 & 16:48:22.043(7) & {-46:11:15.75}$^{0.19}_{0.2}$ & 55395 & 6.0621606076(2) & -87.220(5) & 125 & 10.4 & 54220-58012 \\
J1649--4653 & 16:49:24.61(11) & -46:53:09.3(2) & 55360 & 1.79521547256(18) & -15.98143(2) & 125 & 10.4 & 54220-58012 \\
J1650--4921 & 16:50:35.109(17) & -49:21:03.76(3) & 55599 & 6.393872581394(2) & -7.43411(5) & 112 & 9.5 & 54548-58012 \\
J1702--4306 & 17:02:27.36(2) & -43:06:45.1(4) & 55560 & 4.64018878013(2) & -21.05720(2) & 102 & 9.6 & 54505-58012 \\
J1715--3903 & 17:15:14.08(4) & {-39:02:57.13}$^{0.06}_{0.12}$ & 55370 & 3.5907423095(9) & -48.2784(13) & 128 & 10.4 & 54220-58012 \\
J1722--3712 & 17:22:59.21(4) & {-37:12:04.51}$^{0.06}_{0.09}$ & 55362 & 4.2340633683(7) & -19.4742(11) & 131 & 10.4 & 54220-58012 \\
J1723--3659 & 17:23:07.58(17) & -36:59:14.2(8) & 55384 & 4.93279317887(3) & -19.5353(5) & 128 & 10.4 & 54220-58012 \\
J1733--3716 & 17:33:26.760(2) & -37:16:55.19(10) & 55359 & 2.96213003717(4) & -13.19989(9) & 129 & 11.1 & 53971-58012 \\
J1735--3258 & {17:35:56.66}$^{0.61}_{0.09}$ & {-32:58:21.78}$^{0.46}_{0.38}$ & 55355 & 2.84923231813(2) & -21.1107(3) & 89 & 6.7 & 54220-56672 \\
J1738--2955 & 17:38:52.12(2) & {-29:55:57.39}$^{0.22}_{0.15}$ & 55377 & 2.2551713364(2) & -41.7146(12) & 89 & 6.8 & 54220-56709 \\
J1739--2903 & 17:39:34.292(2) & -29:03:02.2(2) & 55385 & 3.09706373618(5) & -7.55345(7) & 135 & 10.4 & 54220-58012 \\
J1739--3023 & 17:39:39.79(4) & {-30:23:12.87}$^{0.18}_{0.08}$ & 55351 & 8.7434194934(13) & -87.1129(2) & 133 & 10.4 & 54220-58012 \\
J1745--3040 & 17:45:56.316(12) & -30:40:22.9(11) & 55276 & 2.721579246363(8) & -7.90460(4) & 219 & 13.6 & 53035-58012 \\
J1801--2154 & {18:01:08.38}$^{0.63}_{0.05}$ & {-21:54:07.51}$^{0.09}_{0.81}$ & 55385 & 2.66452256901(11) & -11.3721(15) & 84 & 6.8 & 54220-56708 \\
J1806--2125 & {18:06:19.59}$^{0.48}_{0.06}$ & {-21:27:55.33}$^{0.98}_{0.48}$ & 55349 & 2.075444041(15) & -50.821(2) & 123 & 11.0 & 53968-57992 \\
J1809--1917 & 18:09:43.136(2) & -19:17:38.1(5) & 55366 & 12.0838226201(8) & -372.7882(19) & 130 & 10.4 & 54220-58012 \\
J1815--1738 & 18:15:14.67(19) & {-17:38:06.95}$^{0.32}_{0.57}$ & 55364 & 5.03887545888(10) & -197.4552(11) & 86 & 6.8 & 54220-56708 \\
J1820--1529 & {18:20:41.11}$^{0.47}_{0.07}$ & {-15:29:42.37}$^{0.08}_{0.29}$ & 55373 & 3.000716562(16) & -34.130(4) & 81 & 7.4 & 53968-56671 \\
J1824--1945 & 18:24:00.56(18) & {-19:46:03.47}$^{0.21}_{0.43}$ & 55291 & 5.281575552287(3) & -14.6048(5) & 149 & 10.4 & 54220-58012 \\
J1825--1446 & 18:25:02.96(17) & {-14:46:53.75}$^{0.72}_{0.68}$ & 55314 & 3.5816835827(4) & -29.0816(6) & 132 & 10.4 & 54220-58012 \\
J1828--1057 & 18:28:33.24(10) & -10:57:26.9(7) & 55334 & 4.05954117419(6) & -34.1114(4) & 89 & 6.8 & 54220-56708 \\
J1828--1101 & 18:28:18.8(13) & {-11:01:51.28}$^{0.02}_{0.93}$ & 55356 & 13.877993641(13) & -284.992(2) & 131 & 11.1 & 53951-58012 \\
J1830--1059 & {18:30:47.51}$^{0.11}_{0.1}$ & {-10:59:26.45}$^{0.88}_{0.74}$ & 55372 & 2.4686900068(5) & -36.5201(10) & 154 & 10.4 & 54220-58012 \\
J1832--0827 & 18:32:37.013(2) & -08:27:03.7(12) & 55397 & 1.544817633127(2) & -15.24858(4) & 124 & 10.4 & 54220-58012 \\
J1833--0827 & 18:33:40.268(3) & -08:27:31.6(18) & 55402 & 11.7249580817(4) & -126.1600(8) & 124 & 10.2 & 54268-58012 \\
J1834--0731 & 18:34:15.97(2) & -07:31:05.93(7) & 55376 & 1.94933571114(6) & -22.1210(7) & 85 & 6.7 & 54268-56708 \\
J1835--0944 & 18:35:46.653(6) & -09:44:27.2(4) & 55130 & 6.88006896072(4) & -20.7560(11) & 41 & 3.7 & 54478-55822 \\
J1835--1106 & 18:35:18.41(5) & -11:06:16.1(9) & 55429 & 6.0270868794(10) & -74.7918(16) & 125 & 10.2 & 54268-58012 \\
J1837--0559 & 18:37:23.652(6) & -05:59:28.6(2) & 55470 & 4.97354763055(2) & -8.1858(4) & 115 & 10.2 & 54303-58012 \\
J1838--0453 & 18:38:11.4(12) & {-04:53:25.57}$^{0.32}_{0.83}$ & 55339 & 2.6255980205(5) & -80.1949(3) & 91 & 6.6 & 54306-56708 \\
J1838--0549 & 18:38:38.065(6) & -05:49:12.1(3) & 55473 & 4.249688210732(2) & -60.3601(5) & 81 & 6.6 & 54306-56708 \\
J1839--0321 & 18:39:37.520(8) & -03:21:10.8(3) & 55522 & 4.187917144798(2) & -21.9566(7) & 70 & 6.6 & 54306-56708 \\
J1839--0905 & 18:39:53.46(3) & -9:05:14.1(8) & 54979 & 2.38677780294(7) & -14.8244(10) & 55 & 4.3 & 54268-55822 \\
J1842--0905 & 18:42:22.15(2) & -09:05:30.0(3) & 55392 & 2.90152784474(2) & -8.8183(4) & 126 & 10.2 & 54268-58012 \\
J1843--0355 & 18:43:06.663(8) & -03:55:56.6(3) & 55402 & 7.557780825655(2) & -5.94013(9) & 84 & 7.5 & 53968-56708 \\
J1843--0702 & 18:43:22.439(2) & -07:02:54.6(14) & 55380 & 5.21880961058(15) & -5.81812(2) & 128 & 10.2 & 54268-58012 \\
\hline
\end{tabular}}
\end{adjustbox}
\end{table*}

\begin{table*}
\ContinuedFloat 
\caption{
Observational characteristics of the 85 pulsars described in this paper. The position and spin-down parameters are reported at the mentioned period epoch (PEPOCH) along with the timespan and the MJD range. The 95\% confidence limits for the position and spin-parameters reported here are derived from the preferred model for each pulsar.  The confidence regions are individually stated, if the upper and lower confidence limits are asymmetric (continued).}
\centering
\begin{adjustbox}{angle=0}
\renewcommand{\arraystretch}{1.5}
\resizebox{\textwidth}{!}{
\begin{tabular}{lrrrrrrrrrrr}
\hline
\hline
JName & RAJ & DECJ & PEPOCH & $\nu$ & $\dot{\nu}$ & $N_{ToA}$ & Timespan & MJD Range  \\
 & (h:m:s) & (d:m:s) & & $(s^{-1})$ & $(10^{-14} s^{-2})$ & & (yr) &  \\
\hline
J1844--0538 & 18:44:05.12(2) & -05:38:34.1(14) & 55410 & 3.91076899155(11) & -14.84390(17) & 122 & 10.2 & 54268-58012 \\
J1845--0743 & 18:45:57.1833(4) & -07:43:38.57(2) & 55336 & 9.551586249996(12) & -3.345425(2) & 130 & 10.3 & 54267-58012 \\
J1853--0004 & 18:53:23.027(8) & -00:04:33.4(3) & 55446 & 9.85832573140(2) & -54.1604(5) & 118 & 10.1 & 54306-58012 \\
J1853+0011 & 18:53:29.980(8) & 00:11:30.6(3) & 55163 & 2.513260850307(15) & -21.17846(12) & 37 & 3.4 & 54597-55822 \\
\hline
\end{tabular}}
\end{adjustbox}
\end{table*}

\section{Timing analysis} \label{timing analysis}
Establishing a phase coherent solution to the ToAs is an important step in the process of pulsar timing. We know that most of the young pulsars have a strong presence of timing noise and frequent glitches which makes it difficult to produce and maintain phase-connected timing solutions. We use the pulsar-timing code, \textsc{tempo2} (\citealt{tempo2}) to attribute relative pulse numbers to the ToAs and obtain phase connection in the timing residuals. 

We split the timing analysis into 2 steps. The first step involves phase connecting the timing residuals.  The second step involves using the phase connected timing solution in the Bayesian timing package, \textsc{temponest}, to construct a complete timing model with stochastic and additional deterministic parameters. \textsc{temponest} allows us to simultaneously model stochastic and deterministic parameters of interest and marginalize over nuisance parameters that are of no interest to this analysis. For example, in one of the timing models, we fitted the timing noise parameters (white noise and red noise) while simultaneously searching over a wide range of position and spin parameters, while keeping the dispersion measure (DM) fixed. We compute a Bayesian log-evidence value associated with the models for each pulsar to determine which timing model is preferred.

  \begin{figure*}
     \makebox[\linewidth]{
        \includegraphics[width=1.12\linewidth]{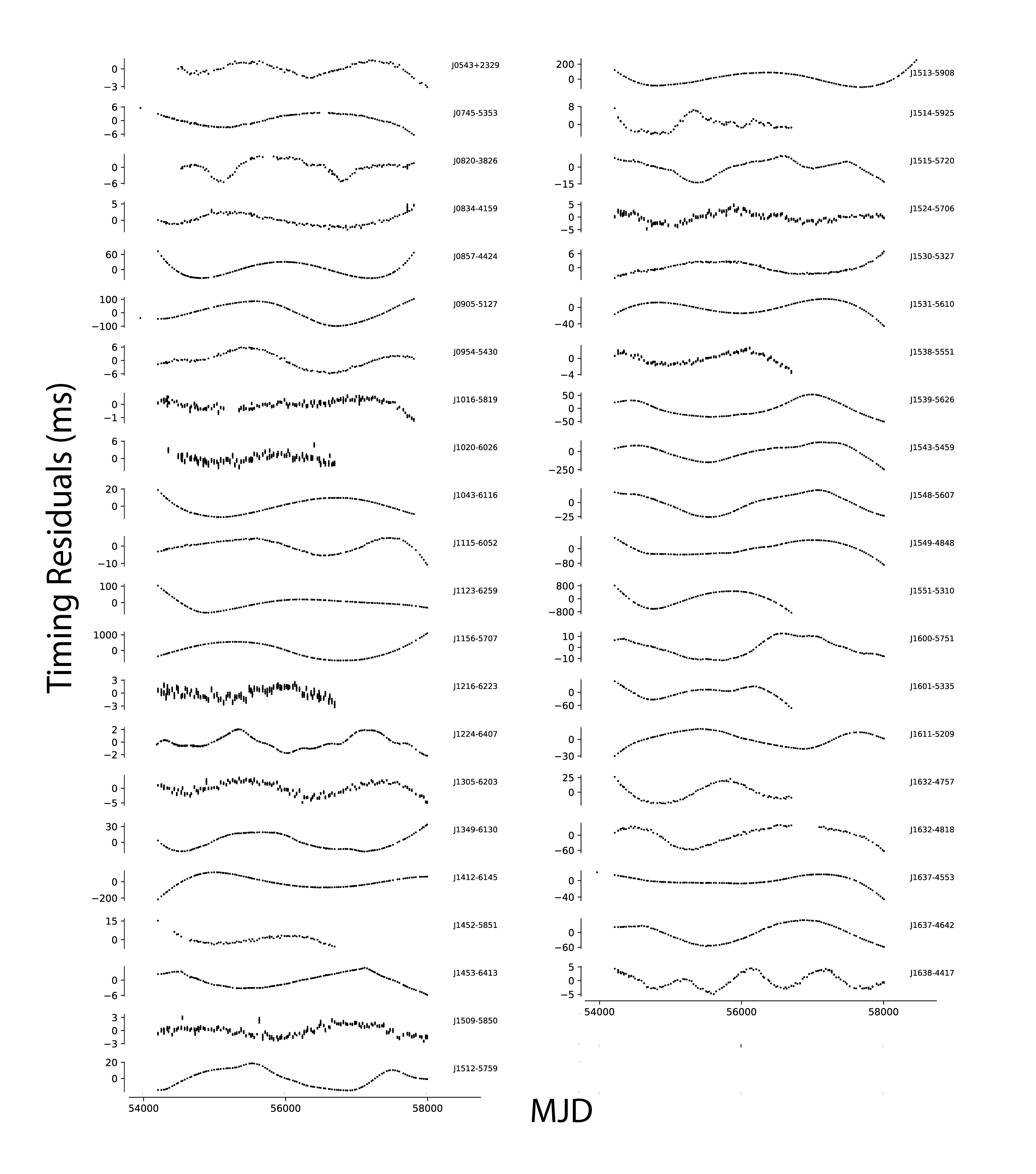}}
     \caption{\label{phaseconnected} Phase-connected timing residuals depicting different levels of timing noise. The timing residuals from the preferred model are shown here, but without removing the contribution of the timing noise.}
  \end{figure*}
  
  \begin{figure*}
  \ContinuedFloat
     \makebox[\linewidth]{
        \includegraphics[width=1.1\linewidth]{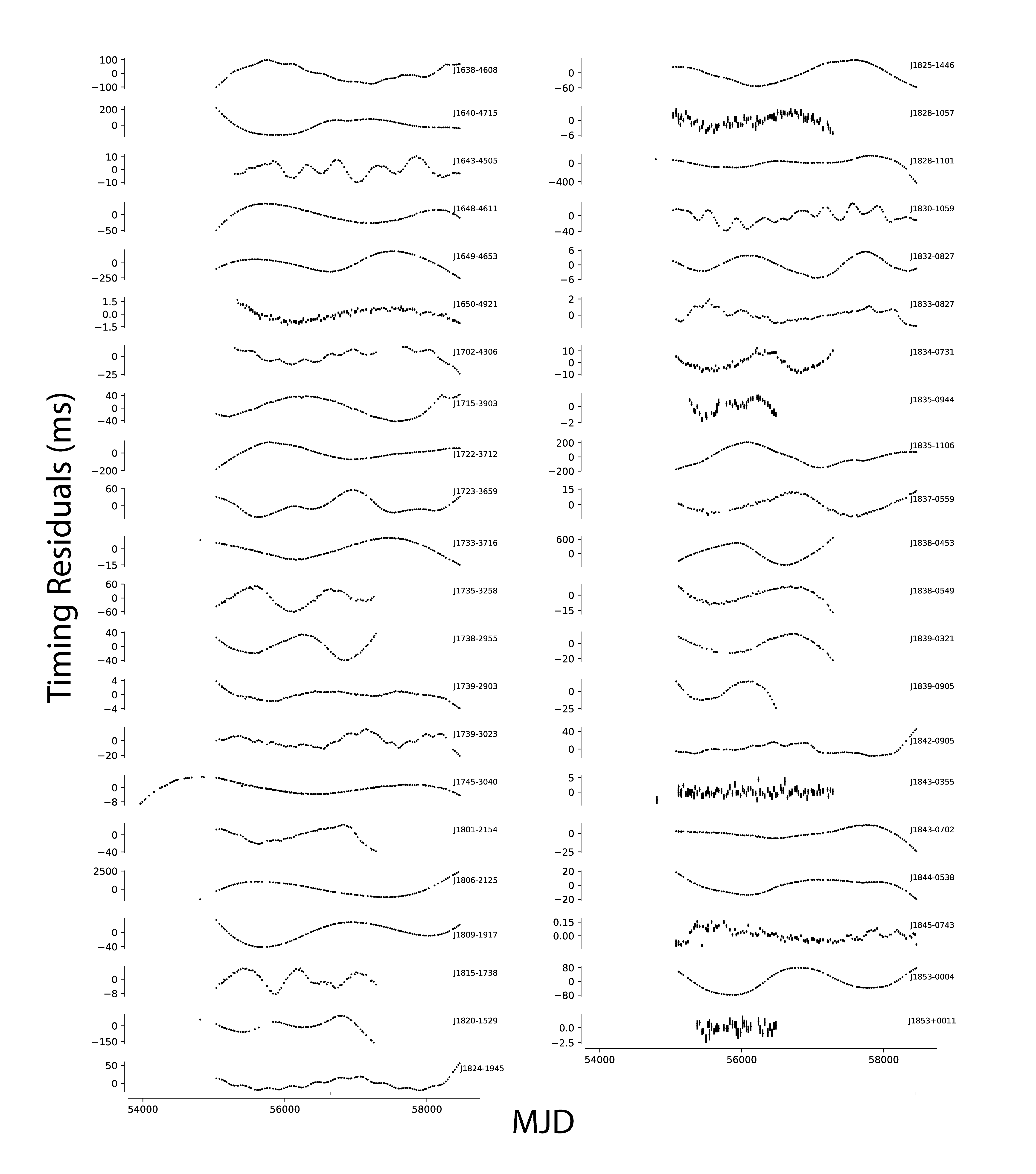}}
     \caption{Phase-connected timing residuals depicting different levels of timing noise. The timing residuals from the preferred model are shown here, but without removing the contribution of the timing noise (continued).}
  \end{figure*}

The ToAs for each pulsar are considered to be a sum of both deterministic and stochastic components: 
  \begin{equation} \label{toa_contribution}
  t_{\rm tot} = t_{\rm det} + t_{\rm sto},
  \end{equation}
The deterministic components in our timing models include various permutations of the pulsar position, spin, proper motion and the spin-down parameters while the stochastic contribution is computed by introducing additional parameters that describe the white and red noise processes. 
The white noise is modelled by adjusting the uncertainty on individual ToAs to be,
  \begin{equation} \label{white_noise}
  \sigma^{2} = F{\sigma_{\rm r}}^{2} + {\sigma_{\rm Q}}^{2},
  \end{equation}
where $F$, referred to as EFAC, is introduced as a free parameter to account for instrumental distortions and ${\sigma_{\rm r}}^{2}$ is the formal uncertainty obtained from ToA fitting. In our analysis we use a global EFAC flag for our 20-cm observations. An additional white noise component (${\sigma_{\rm Q}}^{2}$), commonly referred to as EQUAD, is used to model an additional source of time independent noise measured for each observing system. 
  
%\clearpage
%\includepdf{Plots/Timingplots1.pdf} 
%\includepdf{Plots/Timingplots2.pdf} 

In young pulsars, radio-frequency independent timing noise is the dominant contributing factor towards the red-noise in the ToAs. Many approaches have been taken to improve the parameter estimates by removing some portion of this low-frequency timing noise. \cite{hobbs_whitening} developed a technique to `whiten' the timing residuals using harmonically related sinusoids, that allowed the measurements of proper motions for a large sample of young pulsars using standard timing methods. \cite{coles_whitening} argued that the previously developed ``pre-whitening" methods assumed that the measurements were uncorrelated which resulted in a bias in the parameter estimates. They proposed a new method of improving the timing model fit by using the Cholesky decomposition of the covariance matrix, which described the stochastic processes in the ToAs. They argued that the optimal approach to characterise timing noise, especially those dominated by the presence of strong red noise is to analyze the power spectral density of the pulsar timing residuals. They modelled the timing noise in pulsars using a power-law model to fit for an Amplitude ($A$) and a spectral index estimate ($\beta$). This technique has been used to determine the timing noise parameters and proper motions of millisecond pulsars (\citealt{reardon_timing}). 

%Timing residuals are often correlated and various reasons including improper calibration, failure to properly model the variations due to the ISM and intrinsic timing noise in the pulsar have been identified as contributing to this correlation. Systematic biases and incorrect estimation of both the best-fitting parameters and their uncertainties are the main consequences of neglecting this correlation. 

\cite{tn_haasteren} later developed a joint analysis of the deterministic timing model and the stochastic parameters using a Markov Chain approach and argued that the stationarity of the time-correlated residuals breaks down in the fitting process and that failure to account for the covariances between the deterministic and stochastic parameters leads to incorrect estimation of the uncertainties in the spectral estimates, especially for quadratic spin-down parameters. However, \cite{lentati_2013} pointed out that because the parameter space changes with the linearisation of the timing model, it becomes difficult to perform model selection with the approach in \cite{tn_haasteren}. 

In our analysis we do not search for DM variations and fix the value for the DM in all the models. This is justified as \cite{petroff_dm} found only upper limits to DM variations in the pulsars under consideration here. We model the timing noise as a power-law power spectrum characterised with a red-noise amplitude ($A_{\rm red}$) and a spectral index ($\beta$):
  \begin{equation} \label{red_noise}
  P_{\rm r}(f) = \frac{A_{\rm red}^2}{12\pi^2} \left( \frac{f}{f_{\rm yr}}\right )^{\rm -\beta} ,
  \end{equation}
where $f_{yr}$ is a reference frequency of 1 cycle per year and $A_{\rm red}$ is in units of yr$^{3/2}$.

Motivated by the observations of quasiperiodic timing noise observed in many pulsars, we also model the timing noise as a cut-off power law as described by:
  \begin{equation} \label{cutoff-pl}
  P_{\rm r,CF}(f) = \frac {A (f_{\rm c}/f_{\rm yr})^{\rm -\beta}}{[1+(f/f_{\rm c})^{\rm -\beta/2}]^2},
  \end{equation}
where $f_{\rm c}$ is the corner frequency and $A$ is $(A_{\rm red}^2/12\pi^2)$. We also consider the fact that in young pulsars, the measured timing noise spectral index tends to be steeper as compared to the rest of the pulsar population, with measured values of $\beta \sim 9$ (\citealt{steepRNyoungpulsar}) and so we include low-frequency components with frequencies $f < 1/{\rm T}_{\rm span}$ to model the lowest frequency timing noise. 

A systematic search for periodic modulations in the ToAs is also conducted. We search for harmonic modulations by fitting for a sinusoid with an arbitrary phase, frequency and amplitude and compare the Bayes factor of this model with the others. We also simultaneously model the stochastic parameters with the proper motion parameters to obtain a more robust estimation of the transverse velocity of the pulsar. 

Finally, we search for a braking index, which is caused due to the deceleration of the spin-down rate due to the associated decrease in the magnetic torque. For young pulsars, this braking introduces a measurable second derivative of the spin frequency, 
  \begin{equation} \label{secondspinfreq}
  \ddot{\nu}_{b} = n\frac{\dot{\nu}^{2}}{\nu},
  \end{equation}
and potentially even a third frequency derivative, 
  \begin{equation} \label{thirdspinfreq}
  \dddot{\nu}_{b} = n(2n-1)\frac{\dot{\nu}^{3}}{\nu^{2}}.
  \end{equation}
Analysing the braking indices from a large sample of young pulsars offers a window into the various processes that govern the pulsar spin down. Pulsar braking is a deterministic process and is manifested as low-frequency structures in the ToAs. 

\begin{table}
\caption{\label{tab:prior_ranges}
Prior ranges for the various stochastic and deterministic parameters used in the timing models. $\Delta_{\rm param}$ is the uncertainty on a \textit{parameter} from the initial \textsc{tempo2} fitting.}
\centering
\renewcommand{\arraystretch}{1.5}
\resizebox{\columnwidth}{!}{
\begin{tabular}{lrrrrrrrrrrr}
\hline
\hline
Parameter & Prior range & Type  \\
& &  \\
\hline
Red noise amplitude ($A_{\rm red}$) & (-20,-5) & Log-uniform \\ 
Red noise slope ($\beta$) & (0,20) & Log-uniform \\
EFAC & (-1,1.2) & Log-Uniform \\
EQUAD & (-10,-3) & Log-uniform \\ 
Corner frequency ($f_{\rm c}$) & (0.01/$T_{\rm span}$,10/$T_{\rm span}$) & Log-uniform \\
Low frequency cut-off (LFC) & (-1,0) & Log-uniform \\
Sinusoid amplitude & (-10,0) & Log-uniform \\ 
Sinusoid phase & (0,2$\pi$) & Uniform \\
Log-sinusoid frequency & (1/$T_{\rm span}$,$100/T_{\rm span}$) & Log-uniform \\ 
Proper motion & $\pm$ 1000 mas/yr & Uniform \\ 
RAJ, DECJ, $\nu$, $\dot{\nu}$, $\ddot{\nu}$ & $\pm$ 10000 $\times\Delta_{\rm param}$ & Uniform \\ 
\hline
\end{tabular}}
\end{table}

\subsection{The Bayesian Inference Method}
At the heart of all Bayesian analysis is the Bayes' theorem, which for a given set of parameters $\Theta$ in a model $H$, given data $D$, can be written as:
   \begin{equation}
   \mathrm{Pr}(\Theta \mid D, H) = \frac{\mathrm{Pr}(D\mid \Theta, H)\mathrm{Pr}(\Theta \mid H)}{\mathrm{Pr}(D \mid H)},
   \end{equation}
where
    \begin{itemize}
    \item $\mathrm{Pr}(\Theta \mid D, H) \equiv \mathrm{Pr}(\Theta)$ is the posterior probability distribution of the parameters,
    \item $\mathrm{Pr}(D\mid \Theta, H) \equiv L(\Theta)$ is the likelihood of a particular model,
    \item  $\mathrm{Pr}(\Theta \mid H) \equiv \pi(\Theta)$ is the prior probability distribution of the parameters,
    \item and $\mathrm{Pr}(D \mid H) \equiv Z$ is the Bayesian evidence.
    \end{itemize}

The way we discriminate one model over the other is by considering the evidence ($Z$) which is the factor required to normalize the posterior over $\Theta$, 
   \begin{equation}
   Z = \int L(\Theta)\pi(\Theta) \mathrm{d}^n\Theta,
   \end{equation}
where $n$ is the dimensionality of the parameter space and the ``odds ratio'', $R$,
   \begin{equation}
   R = \frac{Z_{1}}{Z_{2}}\frac{Pr(H_{1})}{Pr(H_{0})},
   \end{equation}
where $\frac{Pr(H_{1})}{Pr(H_{0})}$ is the \textit{a priori} probability ratio for the two models. 
   
Assuming the prior probability of the two models is unity, the odds ratio $R$ reduces to the Bayes factor which is then the probability of one model compared to the other. Since in our analysis we compute the log-evidence, the log Bayes factor is then simply the difference of the log-evidences for the two models. A model is preferred if the log Bayes factor is greater than 5. This states that with equal prior odds, we can expect there to be a $1e^{-5}$ chance, (i.e, 1 in 150) that one hypothesis is true over the other. This is similar to \cite{1909_lentati}, who state that a Bayes factor of $>$ 3 is strong and $>$ 5 is very strong. If multiple models have a Bayes factor greater than 5, we select model A, with a Bayes factor of X, if A is the simpler model and other models have Bayes factors not greater than X+n, where $n=5$ is the threshold. All of these models are computed using the `Bayesian young pulsar timing' pipeline that is cluster-aware and simultaneously processes multiple models for each pulsar. We use $\sim$ 25 different timing models for each pulsar, leading to a total of 2125 models, which were processed in less than 15 hours. The pipeline and the relevant instructions can be found in \url{https://bitbucket.org/aparthas/youngpulsartiming}.

The Bayesian pulsar timing approach is powerful because it allows for the simultaneous modelling of stochastic and deterministic parameters while also allowing for robust model selection based on the principles of Bayesian inference. The unique timing models that we use for each pulsar are:

\begin{itemize}
\item No stochastic parameters (NoSP),
\item Stochastic parameters using a power-law model (PL),
\item Stochastic parameters using a cutoff power-law model (CPL),
\item Proper motion and stochastic parameters (PL+PM),
\item $\ddot{\nu}$ and stochastic parameters (PL+F2),
\item Model with low-frequency components and stochastic parameters (PL+LFC),
\item Model with a single sinusoidal fit and stochastic parameters (PL+SIN).
\end{itemize}

Table \ref{tab:prior_ranges} provides a list of unique parameters used in our different timing models along with their prior ranges. We choose wide prior ranges for the red noise amplitude and spectral index because timing noise in young pulsars is strong and can have a relatively steep spectral index. To perform an unbiased search for the proper motion, our prior distributions range from -1000 mas/yr to +1000 mas/yr. Similarly, to ensure unbiased priors for position, spin and spin-down parameters, the uncertainties of the initial least-squares fit values for each of these parameters are multiplied by $10^{4}$. 

Using various reasonable permutations of these models, we build more sophisticated timing models leading to a total of 25 different models per pulsar. It must be noted that in all of the above models, including the \textit{NoSP} model, the position (RAJ and DECJ) and the spin parameters ($\nu$ and $\dot{\nu}$) are fitted simultaneously with the other relevant model parameters. 

\section{Results} \label{results}
In Table~\ref{obs_char_table} we present the position and spin parameters for the pulsars in our sample with their 95\% credible regions as calculated from the posterior distributions along with their observation timespan. Figure~\ref{phaseconnected} shows the timing residuals from the preferred model, without subtracting the modelled timing noise.  

This is the first time that the timing noise has been consistently modelled using Bayesian inference for a large sample of young pulsars. In Table~\ref{tab:p574_results} we present the preferred timing model, the Bayes factor of that model relative to the base model, which in this case is the model in which the position, spin frequency, spin frequency derivative and a power-law timing noise are fitted for. The Bayes factor is zero if the preferred model is the base model (\textit{PL}). For the first 19 pulsars listed, we report significant detections of $\ddot{\nu}$ and the derived braking index values ($n$) from the preferred model, while for the rest, we report their upper and lower limits as derived from the $\textit{PL+F2}$ model. The braking index is estimated by using equation \ref{brakingindex} on the entire posterior distribution of $\nu$, $\dot{\nu}$ and $\ddot{\nu}$. The values for $A_{\rm red}$ and $\beta$ are derived from the preferred model as stated in the second column.

\begin{figure*}
  \begin{subfigure}[b]{0.5\linewidth}
    \centering
    \includegraphics[width=1.0\linewidth]{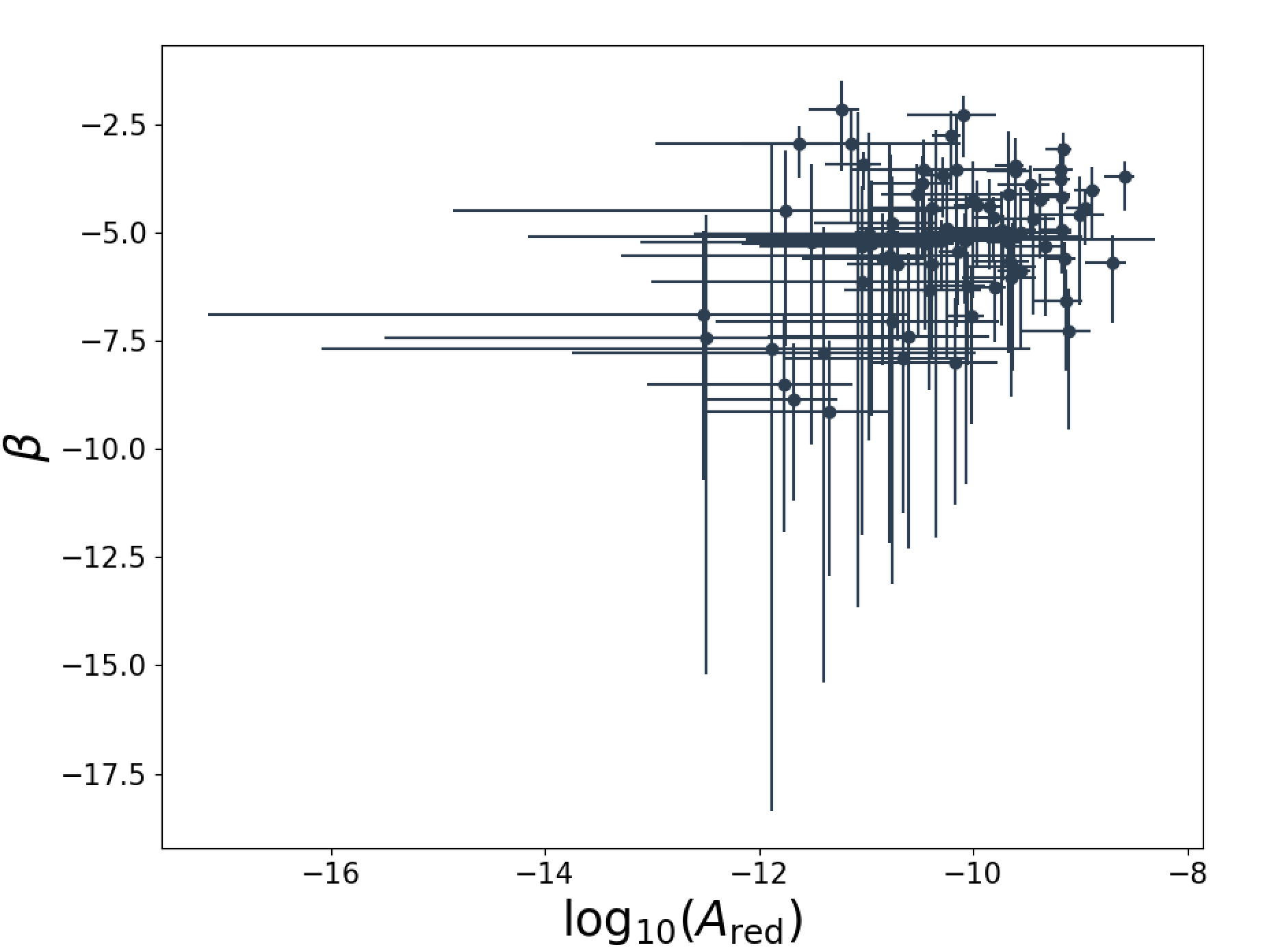} 
    \vspace{4ex}
  \end{subfigure}%% 
  \begin{subfigure}[b]{0.5\linewidth}
    \centering
    \includegraphics[width=1.0\linewidth]{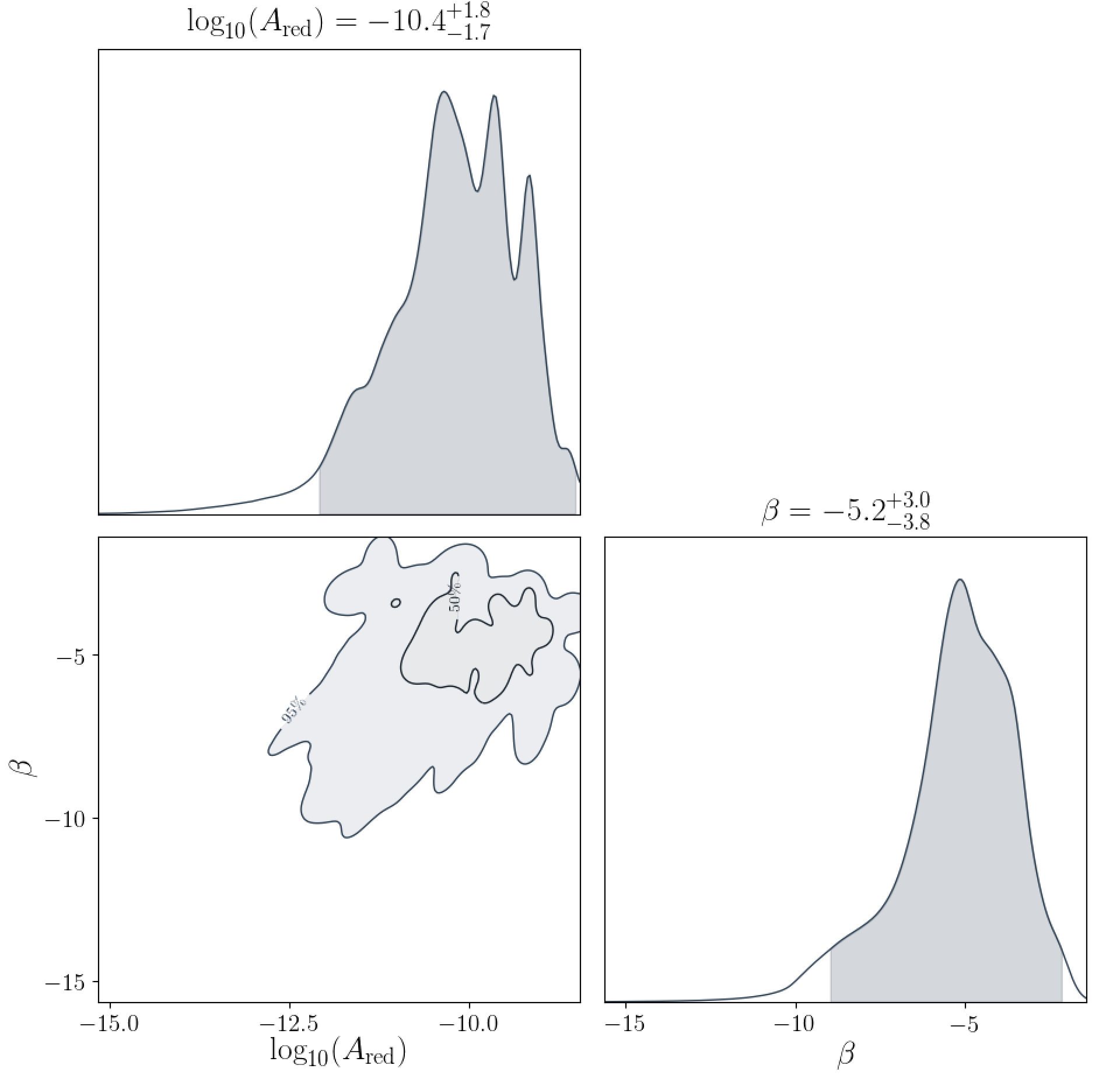} 
    \vspace{4ex}
  \end{subfigure} 
  \caption{\textbf{Left:} The distribution of red noise amplitude against spectral indices for our sample of pulsars for which a power-law timing model is preferred. The error bars are 95\% confidence limits obtained from the preferred model. \textbf{Right:} Posterior distribution of the red noise amplitude and spectral indices from the preferred model for each pulsar are normalized and added together to form an integrated posterior distribution as shown here.
  \label{temponest_post}} 
\end{figure*}

We find that for two pulsars, PSR J1843--0355 and PSR J1853+0011, a model without the timing noise is preferred, while for 58 other pulsars, a model with only the power-law timing noise is strongly preferred. There is marginal to strong evidence for the presence of low-frequency components which are much longer than the data set for five pulsars. We find marginal evidence supporting a cut-off frequency in the power-law timing noise model for PSR J1512--5759. 

A model with a $\ddot{\nu}$ is preferred for 19 pulsars, out of which for three pulsars, the model with low-frequency components is preferred, and for one other pulsar a model with a proper motion is preferred. The braking indices for these pulsars, along with the implications on glitch recovery models and pulsar spin-down are discussed in a second paper (Parthasarathy et al., in prep.). A model with only the proper motion is preferred for PSR J0745--5353. Table~\ref{tab:p574_propermotions} lists the values for the proper motion in right ascension and declination in mas yr$^{-1}$, i.e., $\mu_{\alpha} = \dot{\alpha}\cos\delta$ and $\mu_{\delta}=\dot{\delta}$ and contains the computed transverse velocity using the distance derived from the DM using the electron-density model of \cite{ymw16}. PSR J1702--4306 shows indication for periodic modulation in its ToAs, which is discussed further in Section \ref{modulation_subsec}. It was noted that for PSR J1830--1059, an unpublished glitch was reported\footnote{http://www.jb.man.ac.uk/~pulsar/glitches/gTable.html}, on July 29, 2009 (MJD 55041). For this pulsar a model with a glitch, a $\ddot{\nu}$, and a cut-off power law fit is preferred. It is interesting to note here that we find an evidence for a cut-off frequency for only 2 pulsars out of the 85 in our sample.

Figure~\ref{temponest_post} shows the distribution of $A_{\rm red}$ and $\beta$ extracted from the preferred model for each pulsar in our sample except for the two pulsars for which the cut-off power law model is preferred. The errors shown in the plot are the 2.5\% and 97.5\% confidence limits on both the parameters. The median value for $\log_{\rm 10}(A_{\rm red})$ is $-10.4^{+1.8}_{-1.7}$ yr$^{\rm 3/2}$ and for $\beta$ is $-5.2^{+3.0}_{-3.8}$.  Figure~\ref{temponest_post} also shows the integrated posterior distribution for the timing noise parameters. Contours are plotted for the 50\% and 95\% confidence intervals with the accompanying histograms. 

%  \begin{figure*}
%  \centering
%  \includegraphics[angle=0,width=0.8\textwidth]{Plots/TN_posteriors.pdf}
%  \caption{\label{temponest_post} 
%  The measured red noise amplitude ($A_{\rm red}$) and spectral indices ($\beta$) for our sample of pulsars %extracted from the favoured model, represented by a unique colour, for each pulsar (left) along with the %integrated posterior distribution (right).}
%  \end{figure*}
  
We test the robustness of the timing noise model by comparing them to an independent Bayesian analysis tool, \textsc{Enterprise} \footnote{https://github.com/nanograv/enterprise} (Enhanced Numerical Toolbox Enabling a Robust Pulsar Inference Suite), which is developed for timing noise and gravitational wave analysis in pulsar timing data. With \textsc{Enterprise}, we use a Parallel-Tempering Ensemble Markov Chain Monte Carlo (PTMCMC) sampler. The prior ranges for the noise models are identical and in both the cases the red noise is modelled as a power law. Since \textsc{Enterprise} does not allow for full non-linear sampling of the timing model and only does implicit marginalization over the parameters in the linear perturbation regime, we compare the noise models for the pulsars that prefer the power-law model only. The distributions are similar to that shown in Figure \ref{temponest_post} with median values of the $\log_{\rm 10}(A_{\rm red})$ and $\beta$ being, $-10.4^{+1.8}_{-1.7}$, $-10.3^{+1.6}_{-1.8}$ and $-5.2^{+3.0}_{-3.8}$, $-5.2 \pm 3.3$ using \textsc{temponest} and \textsc{Enterprise} respectively.

%  \begin{figure}
%  \centering
%  \includegraphics[angle=0,width=0.45\textwidth]{enty_temp_plots_vert.jpg}
%  \caption{\label{temponest_enty_comparison}
%  Comparison of red noise parameters between Enterprise and \textsc{temponest} for pulsars that prefer the `onlySP' model.}
%  \end{figure}
 
\begin{table*}
\caption{\label{tab:p574_results} A summary of the preferred timing model, its Bayes factor compared to the base model, and 95\% confidence limits on the timing noise parameters ($A_{\rm red}$ and $\beta$) for each pulsar are reported. The first 19 pulsars listed have a significant detection of $\ddot{\nu}$ and $n$ while for the rest the lower 2.5\% and  upper 97.5\% confidence limits are reported from the \textit{PL+F2} model.}
      
\centering
\renewcommand{\arraystretch}{1.5}
\resizebox{\textwidth}{!}{
\begin{tabular}{lrrrrrrrrrrr}
\hline
\hline
PSR & Best Model & Bayes factor & $\log_{\rm 10}(A_{\rm red})$ & $\beta$ & $\ddot{\nu}$ & $n$  \\
 &  &  & (yr$^{3/2}$)  & & $(10^{-23}s^{-3})$ &  \\
\hline
J0857--4424 & PL+F2 & 171.61 & ${-11.3}^{+1.2}_{-0.6}$ & ${-9.1}^{+3.8}_{-1.6}$ & 3.63(16) & 2890(30) \\
J0954--5430 & PL+F2 & 5.96 & ${-10.4}^{+0.6}_{-0.3}$ & ${-4.4}^{+2.1}_{-0.8}$ & 0.032(8) & 18(9) \\
J1412--6145 & PL+F2 & 29.99 & ${-10.7}^{+1.1}_{-0.6}$ & ${-7.9}^{+3.6}_{-1.6}$ & 0.62(4) & 20(3) \\
J1509--5850 & PL+F2 & 6.54 & ${-11.1}^{+3.1}_{-2.1}$ & ${-5.1}^{+8.6}_{-2.9}$ & 0.12(16) & 11(3) \\
J1513--5908 & PL+F2 & 44.08 & ${-9.7}^{+0.4}_{-0.2}$ & ${-5.7}^{+1.3}_{-0.6}$ & 189.6(2) & 2.82(6) \\
J1524--5706 & PL+F2 & 13.99 & ${-10.2}^{+1.0}_{-0.7}$ & ${-3.6}^{+3.6}_{-1.3}$ & 0.038(2) & 4.2(7) \\
J1531--5610 & PL+F2 & 100.57 & ${-11.8}^{+1.3}_{-0.6}$ & ${-8.5}^{+3.4}_{-1.6}$ & 1.37(2) & 43(1) \\
J1632--4818 & PL+F2 & 18.69 & ${-9.6}^{+0.8}_{-0.5}$ & ${-5.0}^{+2.7}_{-1.1}$ & 0.48(4) & 6(1) \\
J1637--4642 & PL+F2 & 54.34 & ${-9.7}^{+0.6}_{-0.3}$ & ${-4.9}^{+2.2}_{-0.9}$ & 3.2(15) & 34(3) \\
J1643--4505 & PL+F2+LFC & 3.24 & ${-10.1}^{+0.5}_{-0.3}$ & ${-2.3}^{+1.0}_{-0.4}$ & 0.11(2) & 15(6) \\
J1648--4611 & PL+F2 & 13.13 & ${-10.4}^{+0.8}_{-0.5}$ & ${-6.3}^{+2.3}_{-0.9}$ & 0.44(8) & 40(10) \\
J1715--3903 & PL+F2 & 4.19 & ${-9.2}^{+0.2}_{-0.1}$ & ${-3.8}^{+1.3}_{-0.6}$ & 0.4(11) & 70(40) \\
J1738--2955 & PL+F2 & 5.37 & ${-9.6}^{+0.5}_{-0.2}$ & ${-5.8}^{+2.4}_{-1.0}$ & -0.5(16) & -70(40) \\
J1806--2125 & PL+F2 & 5.56 & ${-9.1}^{+0.3}_{-0.1}$ & ${-6.6}^{+1.6}_{-0.7}$ & 1.1(4) & 90(60)\\
J1809--1917 & PL+PM+F2 & 94.14 & ${-11.7}^{+1.1}_{-0.6}$ & ${-9.0}^{+3.5}_{-1.4}$ & 2.70(3) & 23.5(6) \\
J1815--1738 & PL+F2+LFC & 3.18 & ${-11.8}^{+3.1}_{-1.5}$ & ${-4.5}^{+3.1}_{-1.4}$ & 0.73(8) & 9(3) \\
J1824--1945 & PL+F2+LFC & 32.02 & ${-10.9}^{+0.3}_{-0.1}$ & ${-3.4}^{+0.6}_{-0.3}$ & 0.05(2) & 120(20) \\
J1830--1059 & CPL+F2 & 19.55 & ${-8.5}^{+0.3}_{-0.1}$ & ${-13.6}^{+6.2}_{-2.8}$ & 0.16(19) & 31(7) \\
J1833--0827 & PL+F2 & 15.98 & ${-10.2}^{+0.2}_{-0.1}$ & ${-2.8}^{+1.2}_{-0.6}$ & -0.19(13) & -15(2) \\
\hline
J0543+2329 & PL & -\-- & ${-10.5}^{+0.4}_{-0.2}$ & ${-3.6}^{+1.8}_{-0.7}$ & (-0.07,0.01) & (-2,10) \\
J0745--5353 & PL+PM & 20.13 & ${-10.5}^{+0.5}_{-0.3}$ & ${-3.9}^{+1.7}_{-0.6}$ & (-0.01,0.02) & (-140,680) \\
J0820--3826 & PL+LFC & 6.15 & ${-11.1}^{+1.8}_{-1.0}$ & ${-3.0}^{+1.9}_{-0.8}$ & (-0.15,0.06) & (-480,600) \\
J0834--4159 & PL & -\-- & ${-10.9}^{+1.2}_{-0.8}$ & ${-5.3}^{+4.0}_{-1.5}$ & (-0.02,0.02) & (-20,40) \\
J0905--5127 & PL & -\-- & ${-9.6}^{+0.2}_{-0.1}$ & ${-5.0}^{+0.8}_{-0.4}$ & (-0.06,0.14) & (-40,160) \\
J1016--5819 & PL & -\-- & ${-11.5}^{+1.6}_{-1.0}$ & ${-5.2}^{+4.7}_{-1.8}$ & (-0.04,0.01) & (-70,260) \\
J1020--6026 & PL & -\-- & ${-11.9}^{+4.2}_{-2.4}$ & ${-7.7}^{+10.7}_{-4.7}$ & (0.01,0.04) & (10,30) \\
J1043--6116 & PL & -\-- & ${-10.7}^{+0.5}_{-0.3}$ & ${-5.7}^{+1.8}_{-0.7}$ & (0.01,0.03) & (10,90) \\
J1115--6052 & PL & -\-- & ${-10.5}^{+0.7}_{-0.4}$ & ${-5.2}^{+2.3}_{-0.9}$ & (0.02,0.03) & (10,170) \\
\hline
\end{tabular}}
\end{table*}

\begin{table*}
\ContinuedFloat
\caption{  A summary of the preferred timing model, its Bayes factor compared to the base model, and 95\% confidence limits on the timing noise parameters ($A_{\rm red}$ and $\beta$) for each pulsar are reported. The first 19 pulsars listed have a significant detection of $\ddot{\nu}$ and $n$ while for the rest the lower 2.5\% and  upper 97.5\% confidence limits are reported from the \textit{PL+F2} model (continued).}
\centering
\renewcommand{\arraystretch}{1.5}
\resizebox{\textwidth}{!}{
\begin{tabular}{lrrrrrrrrrrr}
\hline
\hline
PSR & Best Model & Bayes factor & $\log_{\rm 10}(A_{\rm red})$ & $\beta$ & $\ddot{\nu}$ & $n$  \\
 &  &  & (yr$^{3/2}$)  & & $(10^{-23}s^{-3})$ &  \\
\hline
J1123--6259 & PL & -\-- & ${-10.1}^{+0.3}_{-0.2}$ & ${-6.3}^{+1.8}_{-0.8}$ & (-0.05,0.1) & (-340,1300) \\
J1156--5707 & PL & -\-- & ${-9.1}^{+0.2}_{-0.1}$ & ${-5.6}^{+0.9}_{-0.4}$ & (-0.73,-0.04) & (-250,100) \\
J1216--6223 & PL & -\-- & ${-11.0}^{+1.6}_{-0.9}$ & ${-5.0}^{+4.8}_{-2.3}$ & (-0.01,0.01) & (-30,40) \\
J1224--6407 & PL+LFC & 11.09 & ${-11.6}^{+0.5}_{-0.3}$ & ${-2.9}^{+0.8}_{-0.4}$ & (-0.05,-0.02) & (-200,100) \\
J1305--6203 & PL & -\-- & ${-11.0}^{+2.0}_{-1.3}$ & ${-6.1}^{+5.9}_{-2.2}$ & (0.01,0.02) & (1,20) \\
J1349--6130 & PL & -\-- & ${-9.8}^{+0.2}_{-0.1}$ & ${-4.7}^{+1.0}_{-0.5}$ & (-0.03,0.09) & (-200,1000) \\
J1452--5851 & PL & -\-- & ${-11.4}^{+2.4}_{-1.4}$ & ${-7.8}^{+7.6}_{-2.9}$ & (0.02,0.03) & (5,10) \\
J1453--6413 & PL & -\-- & ${-10.3}^{+0.2}_{-0.1}$ & ${-3.7}^{+1.0}_{-0.4}$ & (-0.01,0.02) & (-70,250) \\
J1512--5759 & CPL & 2.99 & ${-10.0}^{+0.2}_{-0.1}$ & ${-7.0}^{+2.5}_{-1.1}$ & (-0.03,0.18) & (-10,130) \\
J1514--5925 & PL & -\-- & ${-9.6}^{+0.3}_{-0.1}$ & ${-3.6}^{+1.6}_{-0.8}$ & (-0.07,0.26) & (-300,1600) \\
J1515--5720 & PL & -\-- & ${-9.8}^{+0.2}_{-0.1}$ & ${-4.4}^{+1.5}_{-0.6}$ & (-0.02,0.08) & (-140,760) \\
J1530--5327 & PL & -\-- & ${-10.8}^{+0.7}_{-0.4}$ & ${-4.8}^{+2.7}_{-1.1}$ & (-0.02,-0.01) & (-200,70) \\
J1538--5551 & PL & -\-- & ${-10.8}^{+1.5}_{-1.0}$ & ${-5.1}^{+4.8}_{-1.9}$ & (-0.12,0.02) & (-140,100) \\
J1539--5626 & PL & -\-- & ${-9.7}^{+0.2}_{-0.1}$ & ${-5.1}^{+1.0}_{-0.5}$ & (-0.09,0.1) & (-570,1140) \\
J1543--5459 & PL & -\-- & ${-9.2}^{+0.2}_{-0.1}$ & ${-4.9}^{+1.0}_{-0.5}$ & (-0.2,0.21) & (-40,80) \\
J1548--5607 & PL & -\-- & ${-10.1}^{+0.3}_{-0.1}$ & ${-5.2}^{+1.5}_{-0.6}$ & (-0.06,0.08) & (-30,60) \\
J1549--4848 & PL & -\-- & ${-9.7}^{+0.2}_{-0.1}$ & ${-5.2}^{+1.3}_{-0.6}$ & (0.02,0.19) & (30,330) \\
J1551--5310 & PL & -\-- & ${-9.1}^{+0.4}_{-0.2}$ & ${-7.3}^{+2.3}_{-1.0}$ & (0.42,1.43) & (10,50) \\
J1600--5751 & PL & -\-- & ${-10.0}^{+0.2}_{-0.1}$ & ${-4.4}^{+1.4}_{-0.6}$ & (-0.06,0.04) & (-1000,1600) \\
J1601--5335 & PL & -\-- & ${-9.6}^{+0.5}_{-0.2}$ & ${-6.0}^{+2.8}_{-1.2}$ & (-0.21,0.32) & (-10,40) \\
J1611--5209 & PL & -\-- & ${-10.1}^{+0.2}_{-0.1}$ & ${-5.5}^{+1.2}_{-0.5}$ & (-0.09,0.03) & (-200,200) \\
J1632--4757 & PL & -\-- & ${-10.6}^{+1.3}_{-0.8}$ & ${-7.4}^{+4.9}_{-1.9}$ & (-0.04,0.17) & (-20,150) \\
J1637--4553 & PL & -\-- & ${-10.3}^{+0.3}_{-0.1}$ & ${-5.2}^{+1.1}_{-0.5}$ & (0.03,0.13) & (40,300) \\
J1638--4417 & PL & -\-- & ${-10.0}^{+0.4}_{-0.3}$ & ${-4.3}^{+2.3}_{-0.9}$ & (-0.07,0.08) & (-430,1000) \\
J1638--4608 & PL & -\-- & ${-8.9}^{+0.2}_{-0.1}$ & ${-4.0}^{+1.2}_{-0.5}$ & (-0.36,0.23) & (-30,40) \\
J1640--4715 & PL & -\-- & ${-9.3}^{+0.3}_{-0.1}$ & ${-5.3}^{+1.6}_{-0.7}$ & (0.06,0.31) & (45,330) \\
J1649--4653 & PL & -\-- & ${-10.2}^{+0.8}_{-0.4}$ & ${-8.0}^{+3.3}_{-1.5}$ & (-0.1,0.02) & (-75,60) \\
J1650--4921 & PL & -\-- & ${-12.5}^{+3.0}_{-1.8}$ & ${-7.4}^{+7.8}_{-2.9}$ & (0.01,0.02) & (25,100) \\
\hline
\end{tabular}}
\end{table*}

\begin{table*}
\ContinuedFloat
\caption{  A summary of the preferred timing model, its Bayes factor compared to the base model, and 95\% confidence limits on the timing noise parameters ($A_{\rm red}$ and $\beta$) for each pulsar are reported. The first 19 pulsars listed have a significant detection of $\ddot{\nu}$ and $n$ while for the rest the lower 2.5\% and  upper 97.5\% confidence limits are reported from the \textit{PL+F2} model (continued).}
\centering
\renewcommand{\arraystretch}{1.5}
\resizebox{\textwidth}{!}{
\begin{tabular}{lrrrrrrrrrrr}
\hline
\hline
PSR & Best Model & Bayes factor & $\log_{\rm 10}(A_{\rm red})$ & $\beta$ & $\ddot{\nu}$ & $n$  \\
 &  &  & (yr$^{3/2}$)  & & $(10^{-23}s^{-3})$ &  \\
\hline
J1702--4306 & PL+SIN & 7.1 & ${-9.6}^{+0.2}_{-0.1}$ & ${-3.5}^{+0.8}_{-0.3}$ & (-0.05,0.24) & (-50,360) \\
J1722--3712 & PL & -\-- & ${-9.2}^{+0.2}_{-0.1}$ & ${-4.2}^{+0.8}_{-0.4}$ & (-0.28,0.07) & (-320,270) \\
J1723--3659 & PL & -\-- & ${-9.6}^{+0.2}_{-0.1}$ & ${-5.9}^{+1.4}_{-0.6}$ & (-0.27,0.13) & (-340,420) \\
J1733--3716 & PL & -\-- & ${-10.8}^{+0.8}_{-0.4}$ & ${-5.6}^{+2.5}_{-1.0}$ & (-0.01,0.01) & (-20,35) \\
J1735--3258 & PL & -\-- & ${-9.0}^{+0.4}_{-0.2}$ & ${-4.6}^{+2.1}_{-0.9}$ & (-0.85,0.22) & (-540,510) \\
J1739--2903 & PL & -\-- & ${-10.5}^{+0.3}_{-0.2}$ & ${-4.1}^{+1.6}_{-0.7}$ & (0.01,0.02) & (3,85) \\
J1739--3023 & PL & -\-- & ${-9.2}^{+0.2}_{-0.1}$ & ${-3.1}^{+0.8}_{-0.4}$ & (-0.13,0.18) & (-15,40) \\
J1745--3040 & PL & -\-- & ${-11.0}^{+1.0}_{-0.6}$ & ${-5.3}^{+2.3}_{-0.9}$ & (-0.01,0.02) & (-20,30) \\
J1801--2154 & PL & -\-- & ${-9.2}^{+0.3}_{-0.1}$ & ${-3.6}^{+1.4}_{-0.6}$ & (-0.15,0.16) & (-305,720) \\
J1820--1529 & PL+LFC & 3.18 & ${-12.5}^{+4.6}_{-1.9}$ & ${-6.9}^{+3.8}_{-1.9}$ & (-1.26,0.32) & (-320,290) \\
J1825--1446 & PL & -\-- & ${-9.5}^{+0.3}_{-0.2}$ & ${-3.9}^{+1.2}_{-0.4}$ & (-0.09,0.07) & (-40,70) \\
J1828--1057 & PL & -\-- & ${-10.8}^{+2.5}_{-1.6}$ & ${-5.6}^{+6.6}_{-2.6}$ & (0.01,0.03) & (4,15) \\
J1828--1101 & PL & -\-- & ${-8.6}^{+0.2}_{-0.1}$ & ${-3.7}^{+0.8}_{-0.4}$ & (-0.02,2.52) & (-1,60) \\
J1832--0827 & PL & -\-- & ${-10.4}^{+0.3}_{-0.1}$ & ${-5.1}^{+1.4}_{-0.6}$ & (-0.01,0.01) & (-10,10) \\
J1834--0731 & PL & -\-- & ${-9.7}^{+0.9}_{-0.6}$ & ${-4.1}^{+3.7}_{-1.4}$ & (-0.07,0.05) & (-30,50) \\
J1835--0944 & PL & -\-- & ${-10.3}^{+1.0}_{-0.6}$ & ${-5.2}^{+6.9}_{-2.5}$ & (-0.1,0.22) & (-150,640) \\
J1835--1106 & PL & -\-- & ${-9.0}^{+0.2}_{-0.1}$ & ${-4.4}^{+0.9}_{-0.4}$ & (-0.48,0.6) & (-50,120) \\
J1837--0559 & PL & -\-- & ${-10.2}^{+0.8}_{-0.5}$ & ${-4.9}^{+3.0}_{-1.1}$ & (-0.04,0.04) & (-300,620) \\
J1838--0453 & PL & -\-- & ${-8.7}^{+0.3}_{-0.1}$ & ${-5.7}^{+1.4}_{-0.6}$ & (-2.57,-0.37) & (-100,30) \\
J1838--0549 & PL & -\-- & ${-10.8}^{+1.6}_{-1.0}$ & ${-7.1}^{+6.1}_{-2.2}$ & (0.08,0.11) & (10,15) \\
J1839--0321 & PL & -\-- & ${-10.1}^{+2.1}_{-1.8}$ & ${-5.2}^{+5.7}_{-1.1}$ & (0.02,0.17) & (20,200) \\
J1839--0905 & PL & -\-- & ${-9.4}^{+0.4}_{-0.2}$ & ${-4.7}^{+2.2}_{-0.9}$ & (0.19,0.37) & (210,510) \\
J1842--0905 & PL & -\-- & ${-9.4}^{+0.2}_{-0.1}$ & ${-4.3}^{+1.4}_{-0.6}$ & (-0.1,0.09) & (-400,670) \\
J1843--0355 & NoSP & 2.64 & NA & NA & NA & NA \\
J1843--0702 & PL & -\-- & ${-10.4}^{+0.5}_{-0.3}$ & ${-5.3}^{+2.0}_{-0.8}$ & (0.02,0.06) & (40,1350) \\
J1844--0538 & PL & -\-- & ${-10.4}^{+0.5}_{-0.2}$ & ${-5.7}^{+2.2}_{-1.0}$ & (-0.01,0.05) & (-6,150) \\
J1845--0743 & PL & -\-- & ${-11.2}^{+0.3}_{-0.2}$ & ${-2.2}^{+1.4}_{-0.7}$ & (-0.01,0.03) & (-60,85) \\
J1853--0004 & PL & -\-- & ${-9.8}^{+0.2}_{-0.1}$ & ${-6.3}^{+1.3}_{-0.6}$ & (-0.04,0.43) & (-10,230) \\
\hline
\end{tabular}}
\end{table*}

\section{Discussion} \label{discussion}
\subsection{Timing noise} \label{tn_subsec}
Various attempts have been made to quantify timing noise in pulsars. \cite{cordeshelfand_timingnoise} proposed an `activity parameter' ($A$) that measured the timing noise relative to the Crab pulsar, 
   \begin{equation}
   \label{activity}
   A = \log_{\rm 10} \left [\frac{\sigma_{\rm  TN,2}(T)}{\sigma_{\rm TN,2}(T)_{\rm Crab}} \right ],
   \end{equation}
where $\sigma_{\rm TN,2}\,(T)$ is the RMS residual phase from a second order least squares polynomial fit. They found that this parameter is strongly correlated with the characterstic age of the pulsars. 
\cite{arzoumanian_timingnoise} measured the strength of the timing noise ($\Delta_{8}$) after a cubic polynomial fit to the ToAs over a time period ($T_8$), of $10^8$~s and found a strong correlation with the pulsar period derivative, 
   \begin{equation}
   \label{arzoumanian}
   \Delta_{8} = \log_{\rm 10} \left (\frac{|\ddot{\nu}|}{6\nu}T_{8}^{3} \right ).
   \end{equation}
\cite{shannon_timingnoise} argued that statistics based on a cubic fit ($\ddot{\nu}$) underestimates the strength of the timing noise and proposed that the RMS timing noise after a second order fit is a more accurate diagnostic (i.e. they simply use $\sigma_{\rm TN,2}\,(T)$ without the Crab as reference). They also developed a metric ($\sigma_{\rm P}$),
   \begin{equation}
   \label{shannon_tneq}
   \sigma_{\rm P} = C_{2}\nu^{\alpha}|\dot{\nu}|^{\beta}T^{\gamma},
   \end{equation}
%and estimated the best-fit values of $C_{2}$, $\alpha$, $\beta$ and $\gamma$ over the entire pulsar population to be $2.2\pm 0.4$, $-1.5\pm 0.1$, $1.2\pm 0.1$ and $1.8\pm 0.1$ 
which linked the timing noise with the measured pulsar parameters. Using a maximum likelihood approach they determined the coefficients $\alpha$, $\beta$, $\gamma$ and the scaling factor ($C2$) given the pulsar parameters and the time span ($T$). 

We characterise the strength of the timing noise in our pulsars using the equation,
   \begin{equation}
   \label{tn_strength_eq}
   \log_{\rm 10}({\sigma_{\rm TN}}^2) = 2\log_{\rm 10}(A_{\rm red}) + \log_{\rm 10} \left (\frac{T}{1 \rm yr} \right)  \,(\beta-1) \;,
   \end{equation}
where $T$ signifies the time span over which $\ddot{\nu}$ is measured. Previous metrics for timing noise relied upon modelling it as either a second-order or a cubic polynomial which directly affected the measurements of higher order spin-down parameters.  Since we characterize the timing noise as a power-law using the amplitude and spectral index, it allows us to measure an unbiased value for the pulsar spin-down parameters. 

To determine the correlation between different pulsar parameters and the strength of the timing noise ($\sigma_{\rm TN})$, we perform a linear least-squares regression analysis between $\sigma_{\rm TN}$ (with $T=10$~y) and $\sigma_{\rm P}$,
  \begin{equation}
  \label{tn_parametrize}
  %\sigma_{\rm TN} = a\ln(\nu) + b\ln(|\dot{\nu}|)
  \sigma_{\rm P} = \nu^a\,|\dot{\nu}|^b.
  \end{equation}
  
  The correlation coefficient ($r$), is computed in a linear regression analysis. We search over the parameter space spanned by arbitrary scaling coefficients, $a$ and $b$ to find the maximally correlated scaling relationship.

\begin{figure*}
  \centering
  \includegraphics[angle=0,width=1\textwidth]{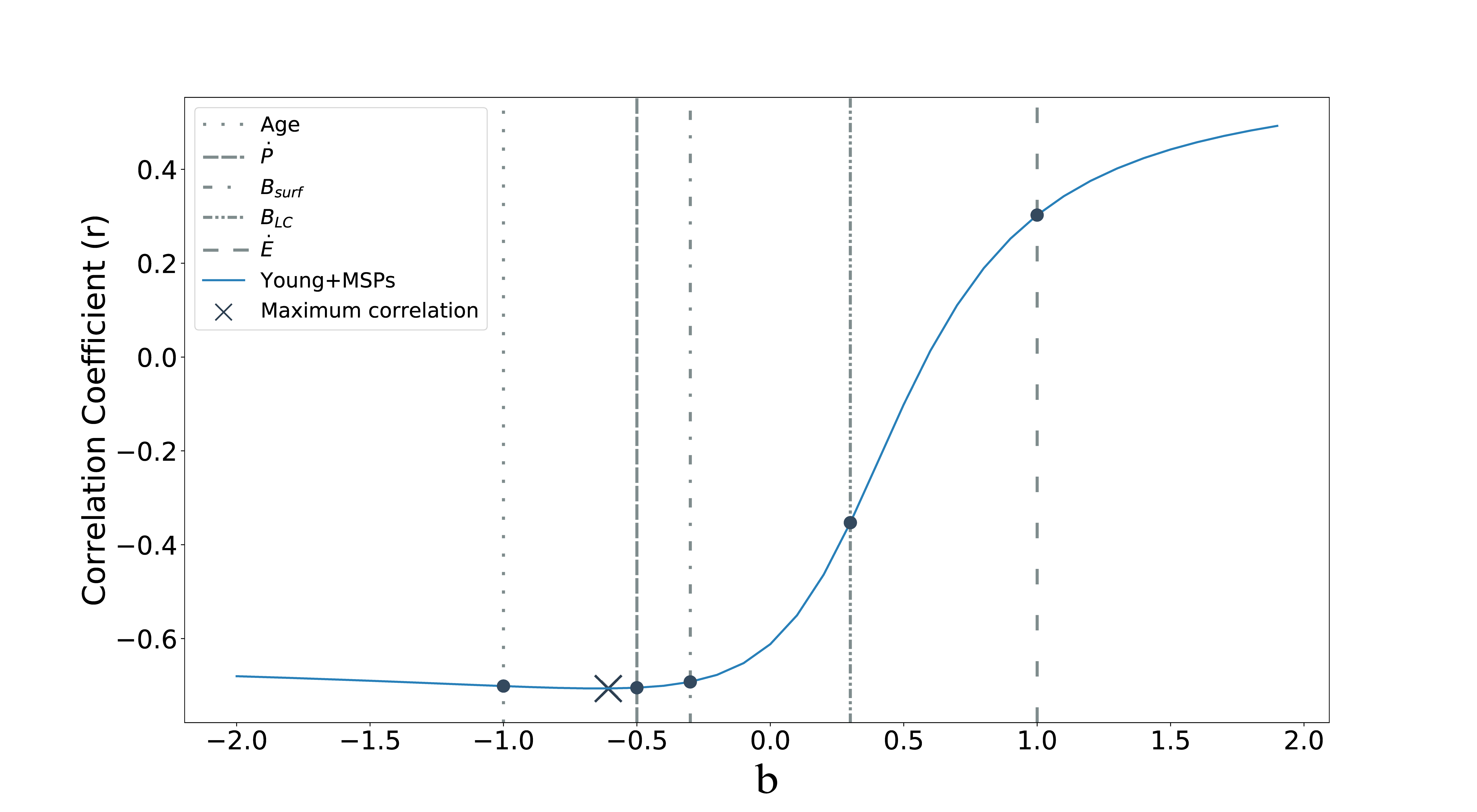}
  \caption{\label{correlation_age} 
  The relationship between the correlation coefficient (r), which measures the strength of the timing noise for various values of $\sigma_{\rm P}$ and $b$, for a fixed value of $a$=1.}
  \end{figure*}

The various pulsar parameters can then be expressed in terms of $\nu$ and $\dot{\nu}$ as:
\begin{itemize}
\item Spin-period derivative: $\nu^{-2}|\dot{\nu}|^1$
\item Spin-down age:  $\nu^{1}|\dot{\nu}|^{-1}$
\item Surface magnetic field strength:  $\nu^{-3/2}|\dot{\nu}|^{1/2}$
\item Magnetic field at the light cylinder:  $\nu^{3/2}|\dot{\nu}|^{1/2}$
\item Rate of loss of rotational kinetic energy:  $\nu^{1}|\dot{\nu}|^1$
\end{itemize}
and are represented in Figure \ref{correlation_age}.

Since the correlation coefficients maintain rotational symmetry in the $(a,b)$ plane and following the discussion in \cite{fabian_spectral}, any combination of $(a,b)$ that has the same ratio will have the same correlation coefficient. For example, the spin-down derivative can be expressed as $\nu^{1}|\dot{\nu}|^{-1/2}$ or $\nu^{-2}|\dot{\nu}|^1$. In our analysis, we set the $a=1$, which results in the various pulsar parameters being expressed as a function of $b$ as shown in Figure \ref{correlation_age}. We then find that the maximum absolute correlation coefficient for $\sigma_{\rm TN}$ occurs at $b = -0.9 \pm 0.2$ for our sample of young pulsars.

This suggests that the timing noise is more closely correlated with spin-period derivative and spin-down age of the pulsar as compared to $\dot{E}$. Analysing the relationship of the timing noise with observing time span, we find no evidence for band-limited timing noise, which would be expected to flatten over longer timing baselines. We compare our results with those of \cite{shannon_timingnoise}, who reported a scaling relation of $\nu^{-0.9\pm0.2}|\dot{\nu}|^{1.0\pm0.05}$, which can also be expressed as $\nu^{1}|\dot{\nu}|^{-1.1\pm0.2}$. We find that our scaling relationships are consistent with those reported by \cite{shannon_timingnoise}.

To test the robustness of this correlation, we also include the timing noise parameters of 8 MSPs from a sample of 49 pulsars from the International Pulsar Timing Array Data release 1 (\citealt{verbiest_ipta}) for which the preferred stochastic model is the spin-noise process \footnote{Uncertainties in the Solar System Barycenter (SSB) have been identified to introduce rednoise signatures in the ToAs of the highest precision MSPs, however, those effects are sub-dominant in the MSP datasets studied here (\citealt{ipta_ssb_arz})} (\citealt{lentati_2016}). The MSPs have a typical observing span of $\sim$10 years and the timing noise is modelled as a power-law process using \textsc{temponest}. 
We find that on adding the MSPs to our sample, we obtain a stronger correlation and the maximum absolute correlation occurs for  $b = -0.6 \pm 0.1$, (Figure \ref{correlation_age}). \cite{tn_haasteren} derive an expression (equation 22 in their paper) for relating the power spectral density to the average RMS in the post-fit timing residuals, which can be used to relate $T^{\gamma}$ in equation \ref{shannon_tneq} to $\beta$ in equation \ref{tn_strength_eq} as $\gamma = \frac{\beta-1}{2}$. From such a relation, we obtain a value of $\gamma$ to be 2 $\pm$ 0.1, consistent with \cite{shannon_timingnoise}. The correlation coefficients obtained for pulsar age, $\dot{E}$ and magnetic field strength are also shown in Figure \ref{correlation_age}. 

\begin{figure*}
  \centering
  \includegraphics[angle=0,width=1\textwidth]{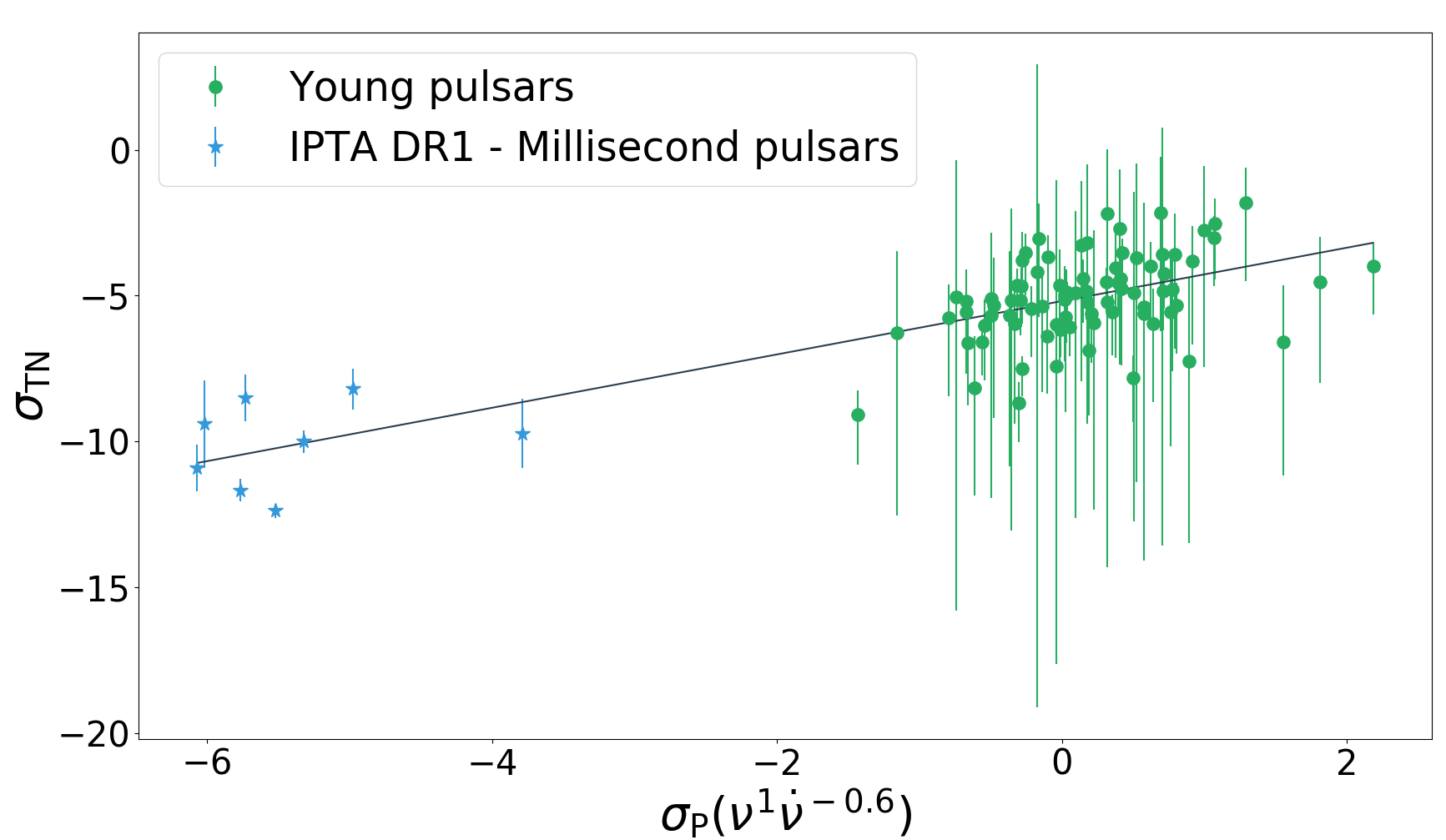}
  \caption{\label{msp_young} 
  Relationship between the timing noise strength and the timing noise metric at the maximally correlated values of $a$ and $b$ for our sample of young pulsars and millisecond pulsars from the International Pulsar Timing Array data release 1 (IPTA DR1) sample.}
  \end{figure*}

Figure \ref{msp_young} shows the correlation between $\sigma_{\rm TN}$ and the timing noise metric ($\sigma_{\rm P}$) for $a = 1$ and $b = -0.6 \pm 0.1$. For the young pulsar sample, the error bars are 95\% confidence limits computed from the measured posterior distributions, while for the MSPs, they are adopted from the $1\sigma$ confidence limits from \cite{lentati_2016}. 
It is evident that the timing noise is stronger in young pulsars as compared to older pulsars (MSPs) in which case, we measure smaller values for the red-noise amplitude and shallower spectral indices. Our parametrization of timing noise from measured values of $A_{\rm red}$ and $\beta$ can be used to predict the relative strength of timing noise in new pulsars given their spin-down parameters. 

\begin{figure}
  \centering
  \includegraphics[angle=0,width=0.5\textwidth]{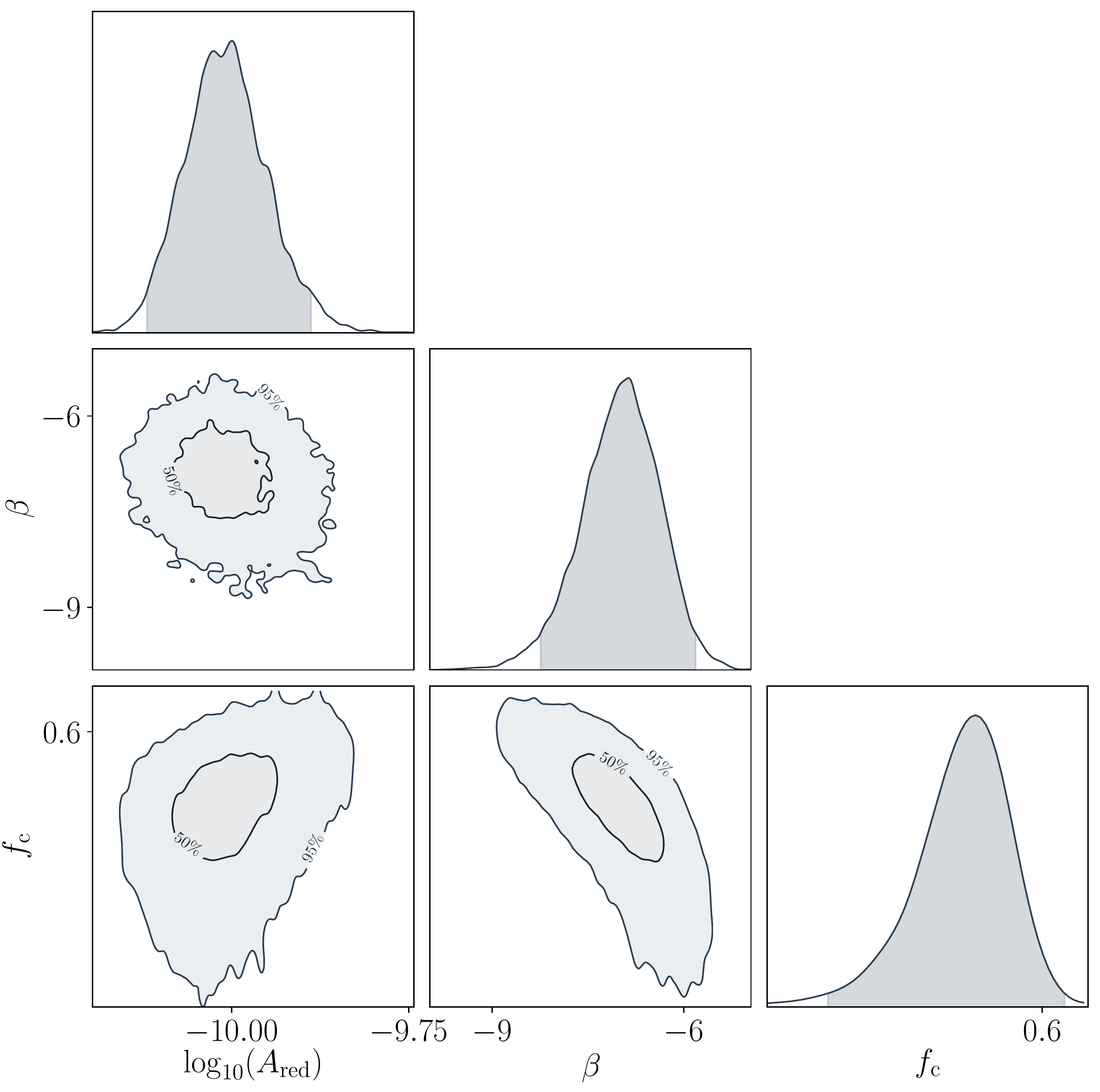}
  \caption{\label{1512_corner} 
  Posterior distribution of the corner frequency parameter along with the timing noise parameters for PSR~J1512--5759. This model is positively preferred with a Bayes factor of 3.23.}
  \end{figure}

\begin{figure}
  \centering
  \includegraphics[angle=0,width=0.5\textwidth]{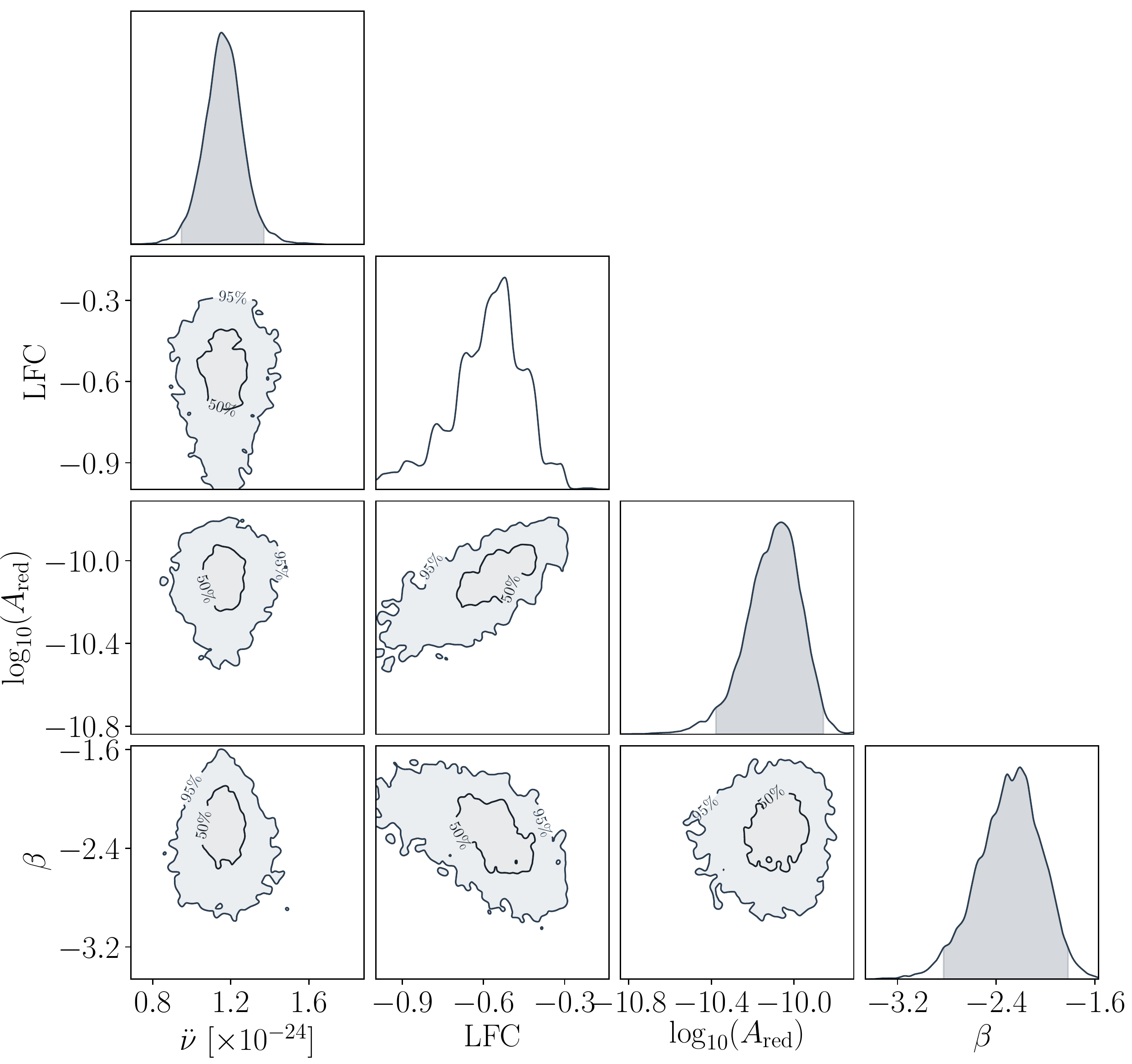}
  \caption{\label{1643_lfc} 
  Posterior distribution of the low-frequency component, a $\ddot{\nu}$ along with the timing noise parameters for PSR~J1643--4505.}
  \end{figure}

We find marginal evidence for the presence of a corner frequency ($f_{\rm c}$) in PSR J1512--5759. The posterior distribution of the corner frequency and the timing noise parameters for this pulsar are shown in Figure \ref{1512_corner}. We find that for five pulsars, a model with a low-frequency component (\textit{LFC}) is preferred. This model implements extra sinusoidal fits at frequencies much longer than the dataset. It is worth noting here that the measurement of low-frequency components is strongly correlated with the amplitude of the red noise (see Figure \ref{1643_lfc}) in the timing residuals. The prospects of detecting signatures at low-frequencies is greater when the red-noise amplitude is larger. This is clearly reflected in the Bayes factors obtained for both PSR J0820--3826 (BF of 6.15) and PSR J1820--1529 (BF of 3.18), which have measured red-noise amplitudes of ${-11.1}^{1.8}_{1.0}$ yr$^{3/2}$ and ${-12.5}^{4.6}_{1.9}$ yr$^{3/2}$ respectively.  For PSR J1643--4505, although the Bayes factor is just 3.24, the measured red-noise amplitude is relatively larger (${-10.1}^{0.5}_{0.3}$ yr$^{3/2}$), thus leading to a relatively well constrained posterior for the LFC parameter as shown in Figure \ref{1643_lfc}. The low Bayes factor can perhaps be attributed to the additional detection of $\ddot{\nu}$. 

\subsection{Proper motions and pulsar velocities} \label{pm_subsec}
\begin{table}
\caption{\label{tab:p574_propermotions}
Proper motions for 2 pulsars reported with their pulsar distance (as estimated from the DM in \protect\cite{ymw16}) and the computed transverse velocities using the proper motion in right ascension ($V_{\rm \alpha T}$) and total ($V_{\rm T}$). The error bars reported are 95\% confidence limits. The epoch for the position is the same as the epoch of the period reported in table \ref{obs_char_table}.} 
\centering
\renewcommand{\arraystretch}{2}
\resizebox{\columnwidth}{!}{
\begin{tabular}{lrrrrrrrrrrr}
\hline
\hline
PSR & $\mu_{\alpha}$ & $\mu_{\delta}$ & $\mu_{\rm tot}$ & Distance &  $V_{\rm \alpha T}$ & $V_{\rm T}$ \\
& (mas/yr) & (mas/yr) & (mas/yr) & (kpc) &(km/s) & (km/s) \\
\hline
%J0745--5353 & ${-62.9}^{10.3}_{9.4}$ & ${51.0}^{10.2}_{10.0}$ & ${81.0}^{10.3}_{9.6}$ & 0.57 & ${-169.9}^{27.8}_{25.3}$ & ${218.8}^{27.7}_{26.0}$ \\ 
%J1809--1917 & ${-19.2}^{6.4}_{6.0}$ & ${53.5}^{87.7}_{86.8}$ & ${56.9}^{82.6}_{81.8}$ & 3.27 & ${-296.8}^{99.7}_{93.2}$ & ${881.4}^{1280.2}_{1267.8}$ \\
%J1824--1945 & ${-133.1}^{94.1}_{98.2}$ & ${-526.9}^{1108.7}_{445.7}$ & ${543.5}^{1075.2}_{432.8}$ & 5.61 & ${-3540.5}^{2502.7}_{2611.7} & ${14452.7}^{28591.3}_{11508.0}$ \\$ \\
%J0745--5353 & ${-63}^{+10}_{-9}$ & ${51}^{+10}_{-9}$ & ${81}^{+9}_{-10}$ & 0.57 & ${-170}^{+28}_{-25}$ & ${219}^{+26}_{-27}$ \\
%J1809--1917 & ${-19}^{+6}_{-5}$ & ${53}^{+88}_{-87}$ & ${58}^{+85}_{-40}$ & 3.27 & ${-297}^{+99}_{-93}$ & ${895}^{+1314}_{-622}$ \\
J0745--5353 & $-60(10)$ & $50(10)$ & $80(10)$ & 0.57 & - & $220(30)$ \\
J1809--1917 & $-19(6)$ & $50(90)$ & $60(90)$ & 3.27 & $-300(100)$ & $900(1300)$ \\
\hline
\end{tabular}}
\end{table}

\begin{figure}   
  \begin{subfigure}[b]{0.4\textwidth}
    \includegraphics[width=\textwidth]{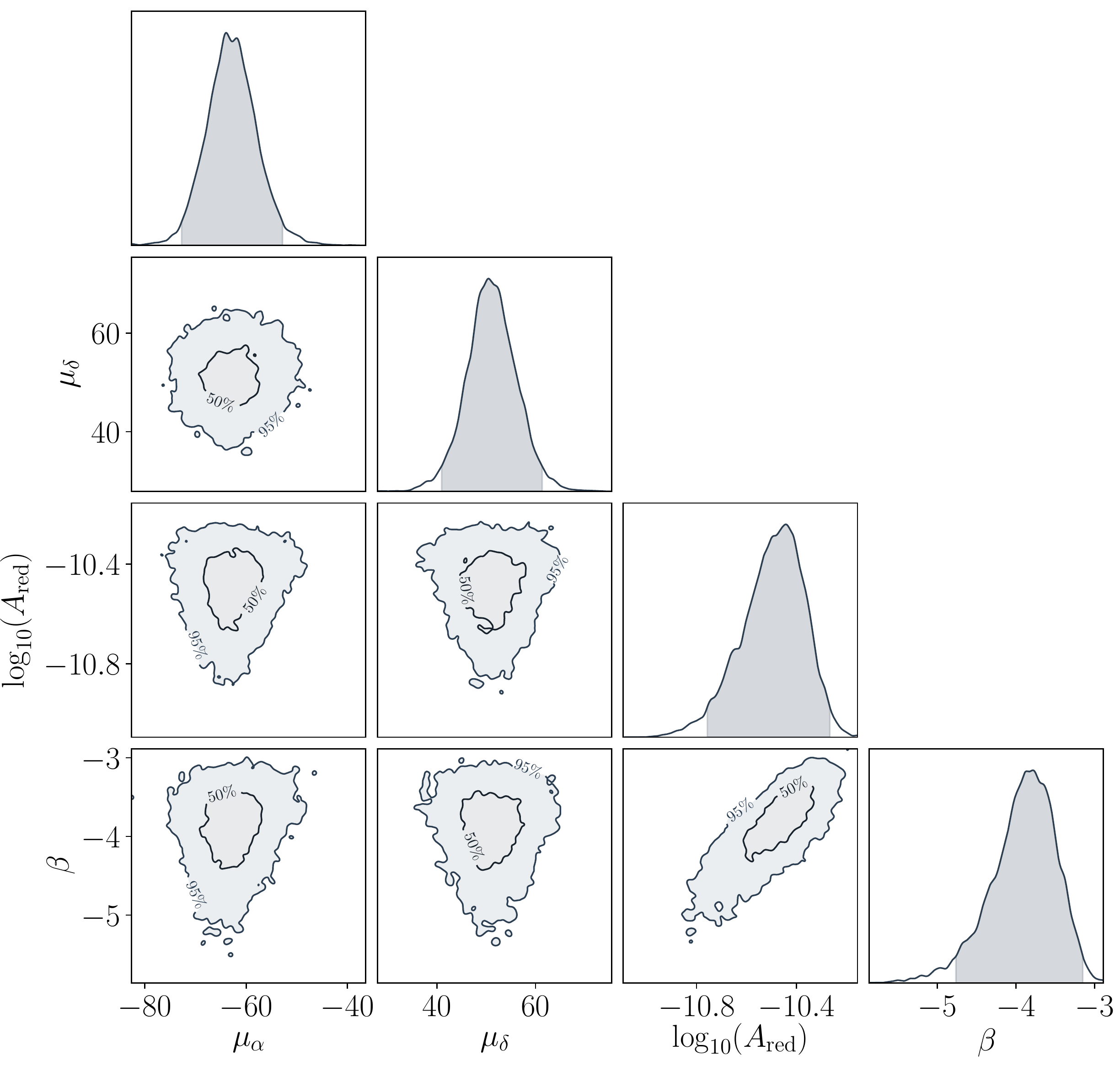}
    \caption{Posterior distribution of the proper motion and the timing noise parameters for PSR J0745--5353.}
    \label{fig:1}
  \end{subfigure}
  \hspace{2ex}
  \begin{subfigure}[b]{0.4\textwidth}
    \includegraphics[width=\textwidth]{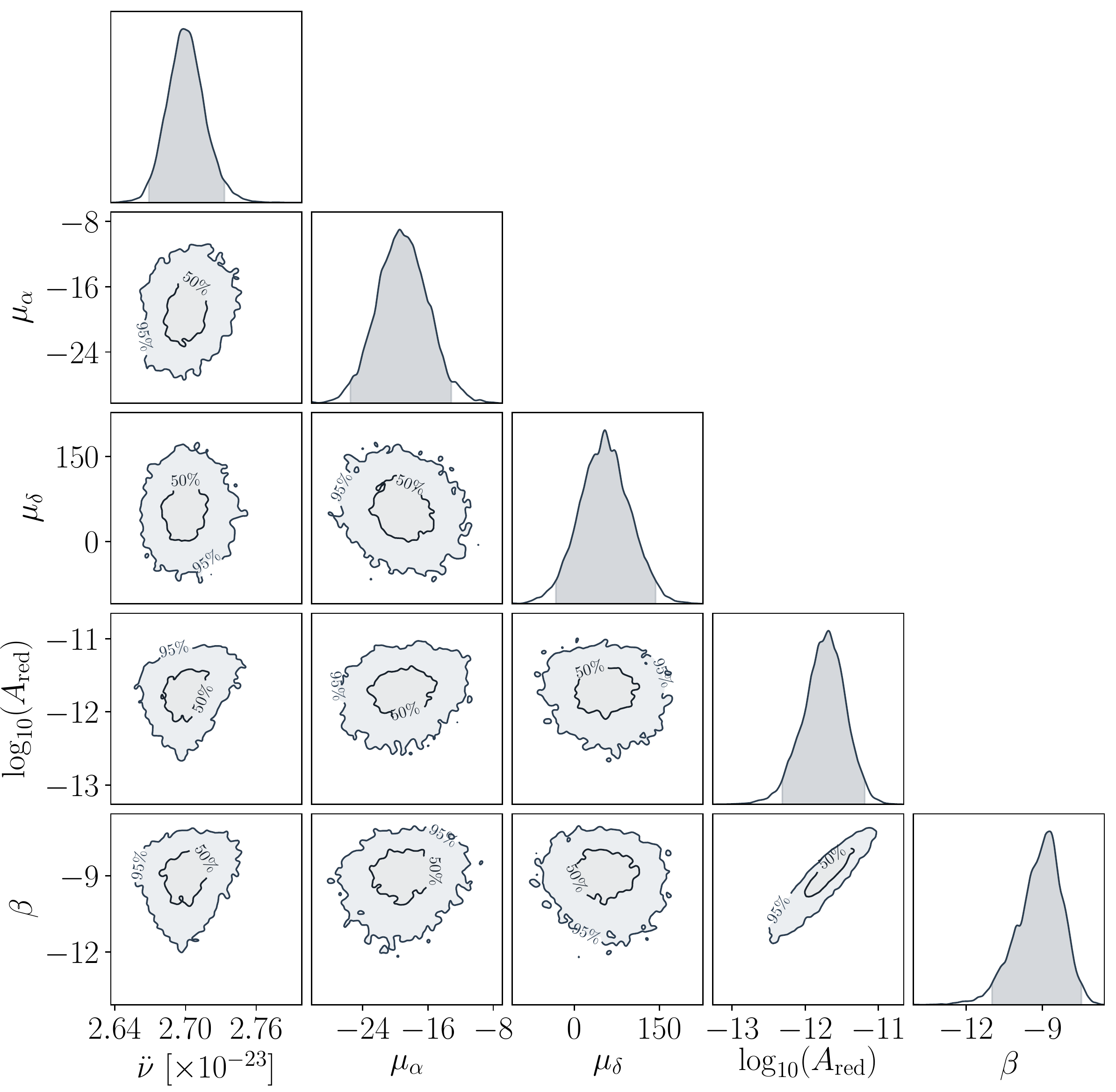}
    \caption{Posterior distribution of the proper motion, $\ddot{\nu}$ and the timing noise parameters for PSR J1809--1917.}
    \label{fig:2}
  \end{subfigure}
  \hspace{2ex}
%  \begin{subfigure}[b]{0.4\textwidth}
%    \includegraphics[width=\textwidth]{posteriors_all/J1824-1945.pdf}
%    \caption{Posterior distribution of the proper motion, $\ddot{\nu}$ and the timing noise %parameters for PSR J1824--1945.}
%    \label{fig:2}
%  \end{subfigure} 
\caption{\label{posterior_pm}}
\end{figure}

For the 2 pulsars listed in Table \ref{tab:p574_propermotions}, the posterior distributions of the proper motions are shown in Figure \ref{posterior_pm}.

PSR J0745--5353 shows a clear detection of a proper motion signature. Assuming a distance of 0.57 kpc, the derived transverse velocity of $220 \pm 30$ km/s is typical of the population of pulsars as a whole. For PSR J1809--1917, we measure a significant proper motion in right ascension, while the proper motion in declination is consistent with zero. The transverse velocity computed from $\mu_{\alpha}$ is $\sim$ 300 kms$^{-1}$, which is reasonable in terms of the transverse velocities for the general pulsar population.

PSR J1745--3040 has a previously reported proper motion from both the frequentist method (\citealt{Zou_PM}), with $\mu_{\alpha}$ of 6 $\pm$ 3 mas/yr, $\mu_{\delta}$ of 4 $\pm$ 26 mas/yr and the Bayesian method (\citealt{pm_bayesian}), with $\mu_{\alpha}$ of 11.9 $\pm$ 16 mas/yr, $\mu_{\delta}$ of 50 $\pm$ 12 mas/yr. In our analysis the proper motion model is marginally better than the power-law model (PL) with a Bayes factor of 2. From this model, we obtain a $\mu_{\alpha}$ of 9.9 $\pm$ 3.5 mas/yr and a $\mu_{\delta}$ of 10.5 $\pm$ 27.6, which are consistent with the previous measurements. PSR J1833--0827 has a previously reported timing proper motion (\citealt{hobbs_propermotion}), but the preferred model in our analysis shows a strong detection of $\ddot{\nu}$. 

There are 2 other pulsars, PSR J1453--6413 (\citealt{bailes_pm}) and J1825--1446 (\citealt{int_pm}) that have a previously reported interferometric proper motions with greater than $3\sigma$ significance. In our analysis, the uncertainties associated with the proper motion measurements are quite large for these pulsars with the preferred models being a power-law model for PSR J1453--6413 and a sinusoidal fitting model for PSR J1825--1446. 

 %\textbf{NOT SURE} We remark that, although only 3 pulsars favour the PM model, there are 7 other pulsars in our sample that have at a $>3\sigma$ detection of $\mu_{\rm tot}$. None of these pulsars have a previously reported proper motion measurement. Both $\mu_\alpha$ and $\mu_\delta$ have a uniform prior range that extends from -600 to +600 mas/yr and since the $\mu_{\rm tot}$ is computed as the quadrature sum of $\mu_\alpha$ and $\mu_\delta$, the probability distribution becomes an exponential. Hence, the probability of a $>3\sigma$ measurement of  $\mu_{\rm tot}$ is $e^{-3}$, or $\sim$4 out of the 85 pulsars in our sample. 
 
Unbiased measurements of proper motion and other such deterministic parameters in pulsars that are strongly contaminated with timing noise strongly underscores the evidence-based model selection that we have employed here. Increasing the timing baselines will help to discover further significant proper motion measurements.

\subsection{Pulsars with planetary companions?}
\label{modulation_subsec}
To search for periodic modulations in our pulsars, we fit for a sinusoid with varied amplitudes, phases and frequencies and compare the evidences to choose the preferred model. Here we comment on five pulsars present in our sample that have been previously studied in the context of periodic signals in their timing residuals.

PSR~J1637--4642 was reported to show marginal evidence for a single sinusoid in \cite{mkerr_precession}. We find that the preferred model for this pulsar is \textit{PL+F2}. In order to further test this, we fitted for a sinusoid simultaneously with $\ddot{\nu}$ but find that this model (PL+F2+SIN) only has a Bayes Factor of 2.9, which does not pass a Bayes factor threshold of 5 over the much simpler model.

PSR~J1825--1446 showed strong evidence for a single sinusoid according to \citealt{mkerr_precession}. We however find that, the \textit{PL+SIN} model does not meet the threshold to be preferred over the \textit{PL} model. The model with a sinusoidal fitting has a Bayes factor of only 1.7. 

For PSR~J1830--1059, we find evidence for a glitch with parameters similar to those in the catalogue and find that the best model is one which includes the glitch, $\ddot{\nu}$ and a cut-off power law model. This pulsar is notable for correlated profile and $\dot{\nu}$ changes (\citealt{brook_profilechanges}, \citealt{mkerr_precession}). \cite{stairs_2019} performed an exhaustive analysis on multi-hour long observations of this pulsar and reported that the pulsar undergoes mode-changing between two stable, extreme profile states. They stated that the observed mode transition rate can perhaps be explained by the chaotic behaviour model as previously suggested by \cite{chaotic_1828}. The detection of a glitch in 2009, further complicates the theoretical models invoking explanations based on pinned vortices inside neutron stars. We conclude that the deviation from a simple power-law, the presence of a glitch and the identified mode changing make this pulsar more complex and demands further investigation. 

PSR~J1638--4608 was reported to show a strong evidence for a single sinusoid fitting in \citealt{mkerr_precession}. Close examination however revealed the presence of 2 new glitches. The amplitudes of these are glitches are very small, in the order of 10$^{-8}$ Hz and 10$^{-9}$ Hz. We find that, after taking the glitches into account, the glitch inclusive model (\textit{GL+SIN}) has a Bayes factor of $\sim$60 as compared to the model with only the stochastic parameters (PL).

It is useful to note here that although the pulsars presented in this analysis were manually selected to not have any identified glitches in the data set, we subsequently found that the two pulsars discussed above had detected glitches. This was missed in the initial manual search owing to the small glitch amplitudes. We decided to retain them in the paper, because for one of the sources, the glitches were unpublished, while for the other, it significantly changed the favoured model. 

For PSR~J1702--4306,  \cite{mkerr_precession} saw strong evidence for a single sinusoid with a projected semi-major axis ($a_{\rm n}$) of 2.9 $\pm$ 0.7 ms and an orbital period ($P_{\rm b}$) of 391 $\pm$ 10 days. In our data, we find that the sinusoidal model is strongly preferred over the \textit{PL} model by a Bayes factor of 7.1. We measure $a_{\rm n}$ to be 2.6 $\pm$ 0.2 ms and $P_{\rm b}$ to be 316 days. It is unclear if these effects are caused due to neutron star precession or due to the presence of a planetary companion, as discussed in \cite{mkerr_precession}.

\section{Conclusions} \label{conclusion}

\begin{comment}
2 - noSP
58 - PL
1 - OnlyPM 
3 - onlyLFC
14 - onlyF2 
1  - onlyCPL
2  - F2 with LFC
1 - onlySIN
2 - PM, F2 (one with LFC)
TODO: 1830 (1)
\end{comment}

We have applied an improved methodology based on Bayesian inference on a large sample of high $\dot{E}$, young pulsars to measure different stochastic and deterministic parameters of interest. We have shown that evidence-based model selection is a powerful technique to disentangle stochastic processes from deterministic ones and to obtain unbiased measurements of pulsar parameters. For each pulsar in our sample, a total of 25 different models were compared and the best model was selected based on a Bayes factor threshold of 5. The power-law model was preferred for 58 pulsars, while we found no evidence of timing noise in two pulsars.  The low-frequency component (\textit{PL+LFC}) model was preferred for five pulsars and in two other pulsars we measure a proper motion signature.  Marginal evidence for the presence of a corner frequency in the power-law was detected in two pulsars. We report two new glitches in PSR J1638--4608 and find evidence for periodic modulation in the ToAs of both PSR J1638--4608 and PSR J1702--4306. We have also compared our timing noise models with an independent Bayesian package, \textsc{enterprise} and obtained consistent results.

We characterize the timing noise as a power-law based on the red-noise amplitude $(A_{\rm red})$ and spectral index ($\beta$) and report that there is a strong correlation between the spin-period derivative of the pulsar and the strength of the timing noise. We develop a metric that can be used to determine the relative strength of the timing noise in any pulsar given its spin-down parameters. On adding MSPs to our sample, we notice that the correlation gets stronger, which is consistent with what is expected. 

Finally, we measure significant $\ddot{\nu}$ measurements for 19 pulsars and also report their braking indices. We discuss the significance of the braking index measurements, their robustness and the effects of glitch recovery models in a subsequent publication. 

\section*{Acknowledgements}
The Parkes radio telescope is part of the Australia Telescope, which is funded by the Commonwealth Government for operation as a National Facility managed by CSIRO. A.P would like to thank Marcus Lower and Daniel Reardon for their comments and ideas. A.P would also like to thank Andrew Jameson for the continued support in installing \textsc{Temponest} on the OzSTAR HPC facility. This work made use of the gSTAR and OzSTAR national HPC facilities. gSTAR is funded by Swinburne and the Australian Government Education Investment Fund. OzSTAR is funded by Swinburne and the National Collaborative Research Infrastructure Strategy (NCRIS). This work is supported through Australian Research Council (ARC) Centre of Excellence CE170100004. A.P. is grateful to CSIRO Astronomy and Space Science for support throughout this work. R.M.S. acknowledges support through ARC grant CE170100004. M.B. acknowledges support through ARC grant FL150100148. Work at NRL is supported by NASA. This work also made use of standard Python packages (\citealt{numpy}, \citealt{scipy}, \citealt{pandas}, \citealt{matplotlib}), Chainconsumer (\citealt{chainconsumer}) and Bokeh (\citealt{bokeh}). 

%%%%%%%%%%%%%%%%%%%%%%%%%%%%%%%%%%%%%%%%%%%%%%%%%%



\bibliographystyle{mnras}
\bibliography{ypt1}

%%%%%%%%%%%%%%%%%%%%%%%%%%%%%%%%%%%%%%%%%%%%%%%%%%

%%%%%%%%%%%%%%%%% APPENDICES %%%%%%%%%%%%%%%%%%%%%

\appendix
\section{Posterior distributions}
\begin{figure*}
\begin{subfigure}[b]{0.4\textwidth}
\includegraphics[width=\textwidth]{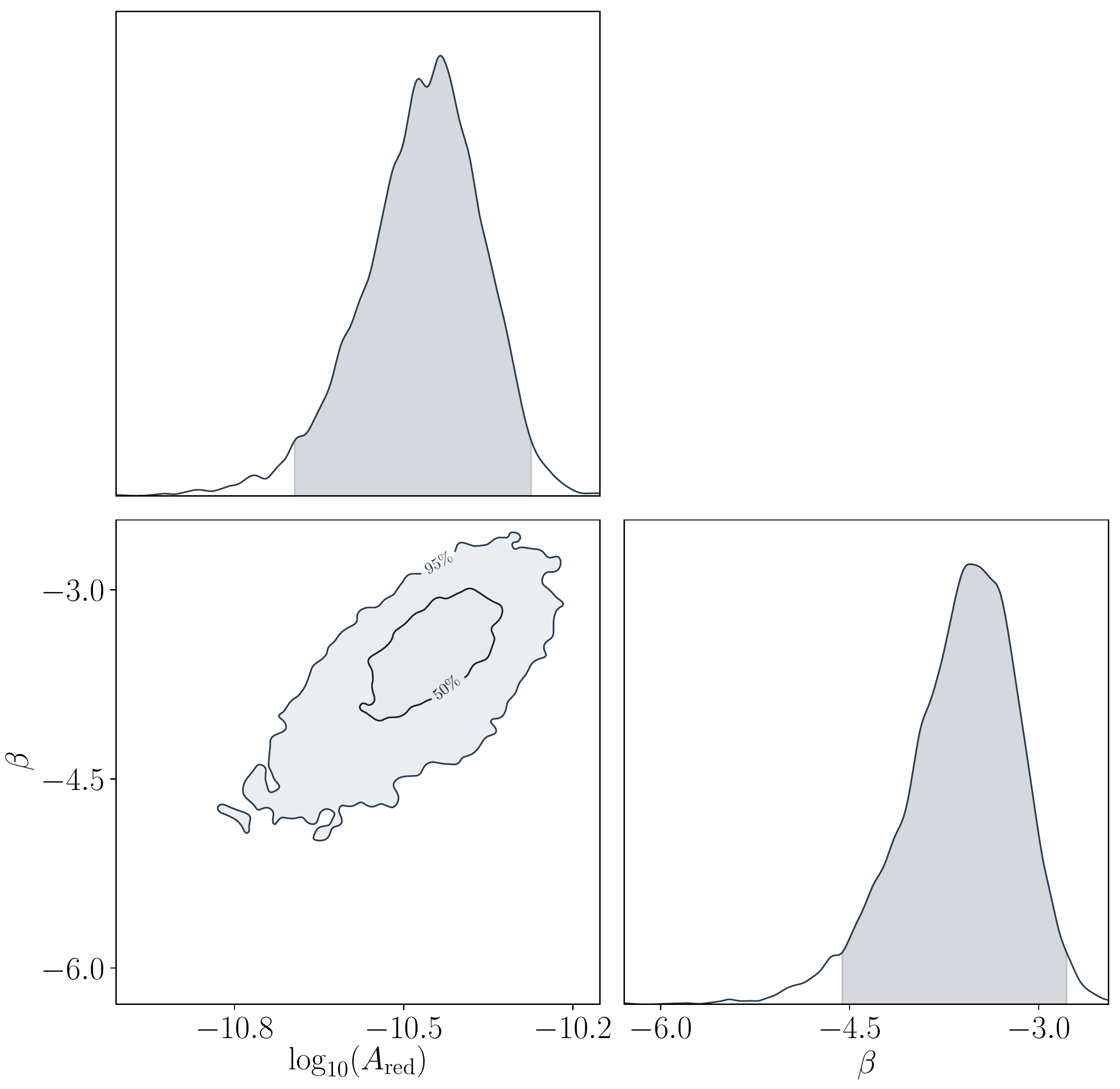}
\caption{Posterior distribution for PSR J0543+2329.}
\end{subfigure}
\hspace{2ex}
\begin{subfigure}[b]{0.4\textwidth}
\includegraphics[width=\textwidth]{posteriors_all/J0745-5353.pdf}
\caption{Posterior distribution for PSR J0745-5353.}
\end{subfigure}
\hspace{2ex}
\begin{subfigure}[b]{0.4\textwidth}
\includegraphics[width=\textwidth]{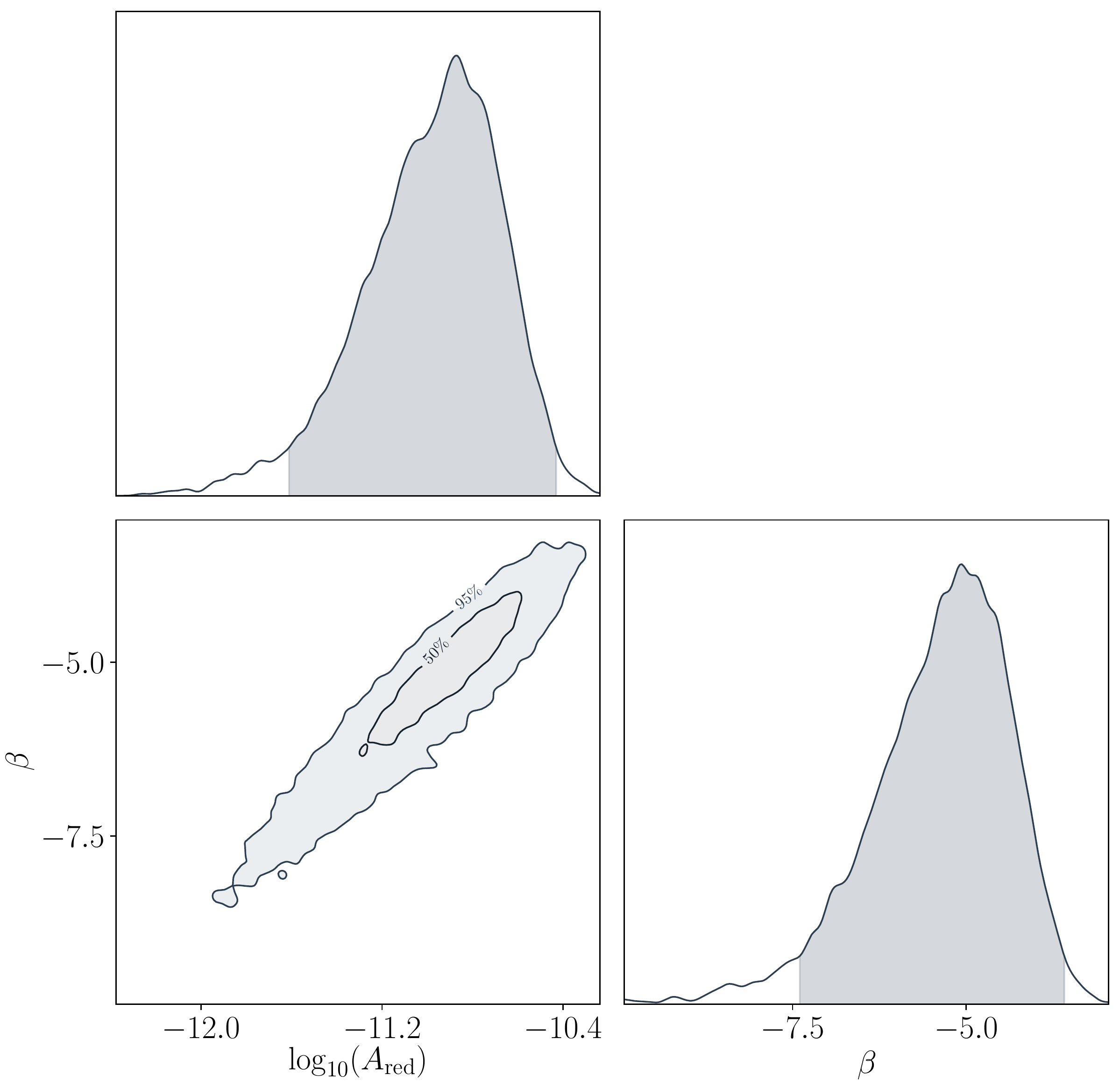}
\caption{Posterior distribution for PSR J0834-4159.}
\end{subfigure}
\hspace{2ex}
\begin{subfigure}[b]{0.4\textwidth}
\includegraphics[width=\textwidth]{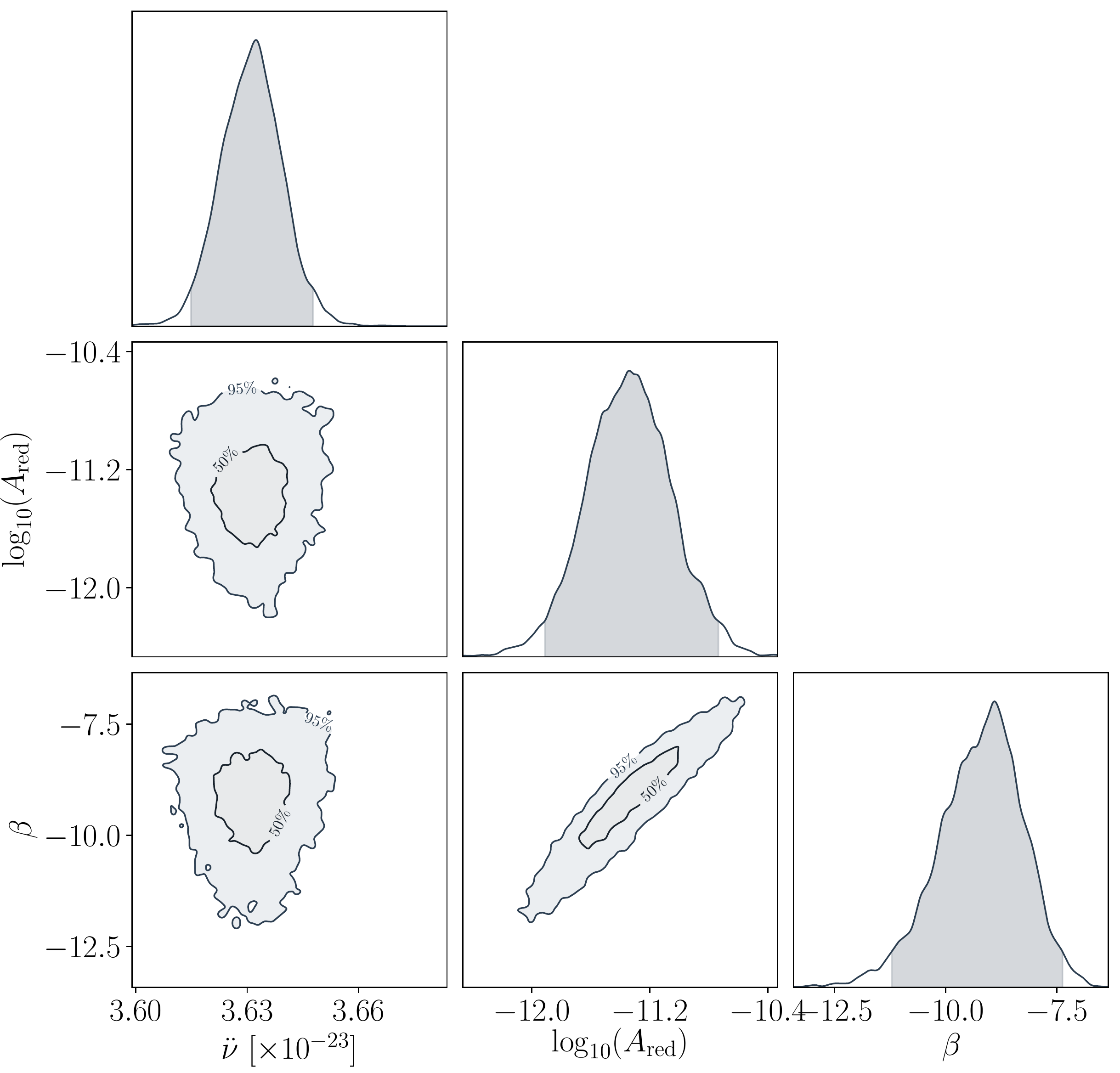}
\caption{Posterior distribution for PSR J0857-4424.}
\end{subfigure}
\hspace{2ex}
\begin{subfigure}[b]{0.4\textwidth}
\includegraphics[width=\textwidth]{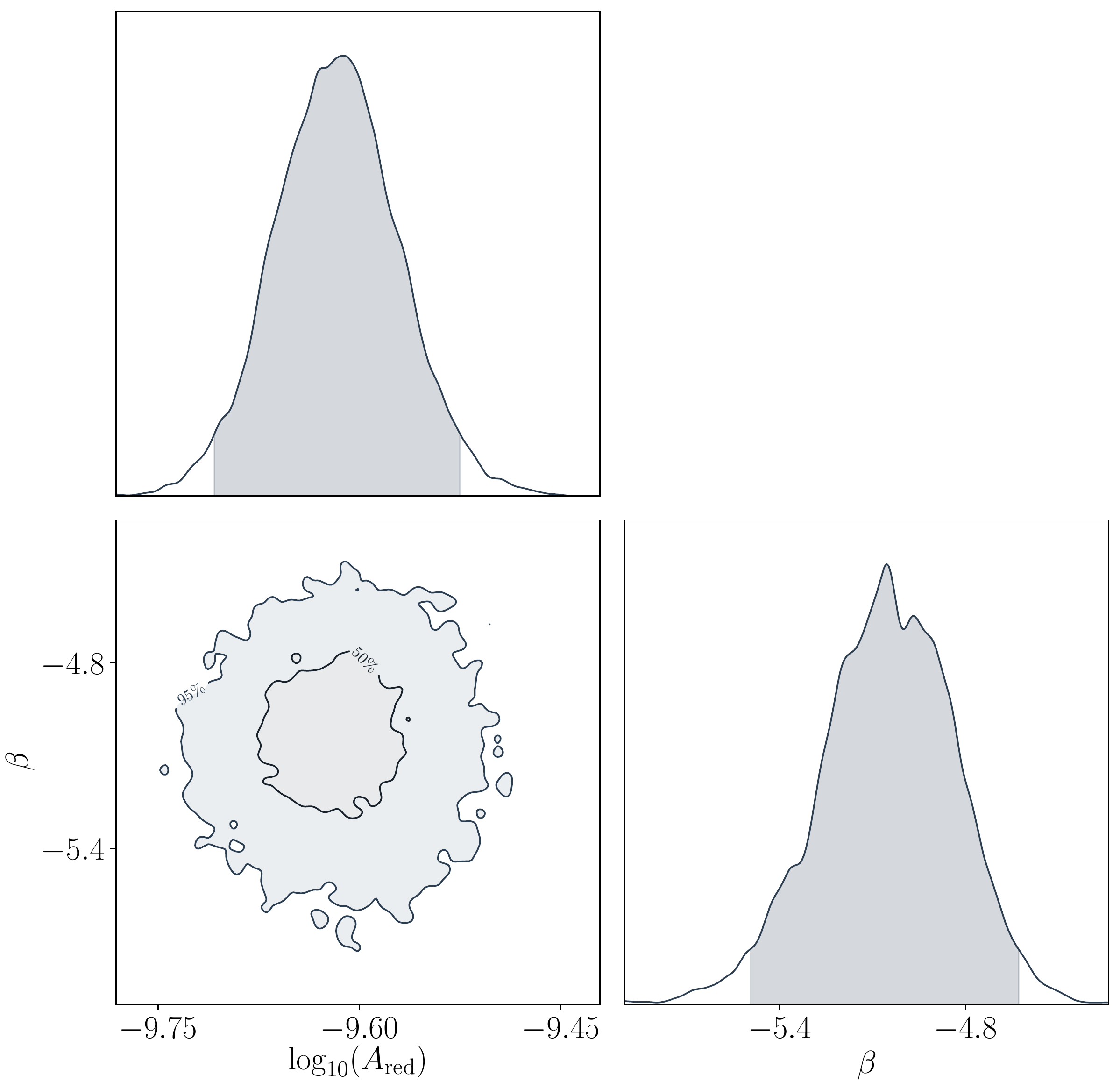}
\caption{Posterior distribution for PSR J0905-5127.}
\end{subfigure}
\hspace{2ex}
\begin{subfigure}[b]{0.4\textwidth}
\includegraphics[width=\textwidth]{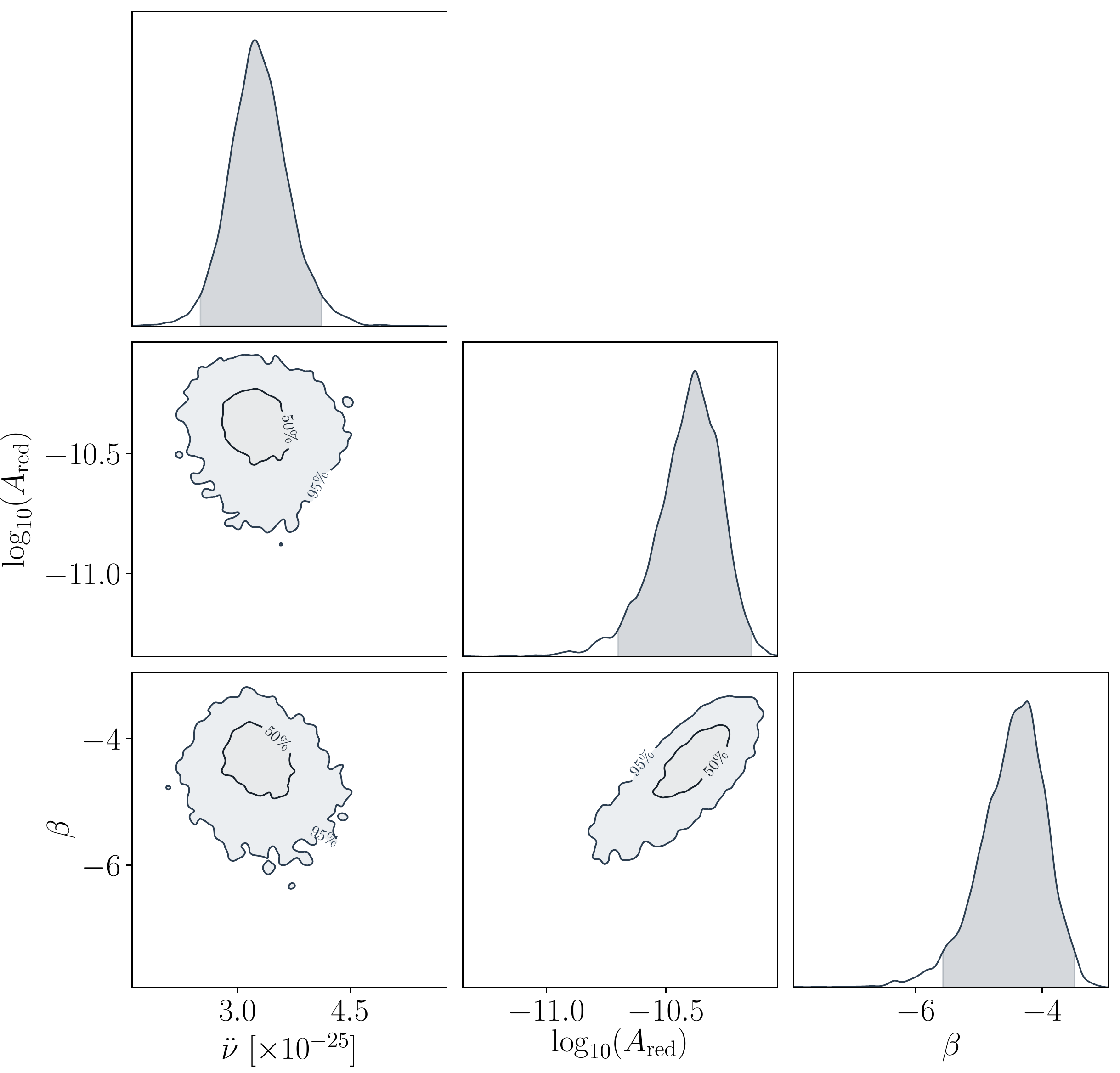}
\caption{Posterior distribution for PSR J0954-5430.}
\end{subfigure}
\caption{\label{posteriors_appendix} Sample posterior distributions of 6 pulsars.}
\end{figure*}

The posterior distributions of the preferred model for 6 pulsars are shown in Figure \ref{posteriors_appendix}  as a sample. Please visit the online repository \url{https://bitbucket.org/aparthas/youngpulsartiming} to view the posterior distributions for all of the 85 pulsars discussed in this paper. 

%%%%%%%%%%%%%%%%%%%%%%%%%%%%%%%%%%%%%%%%%%%%%%%%%%

% Don't change these lines
\bsp	% typesetting comment
\label{lastpage}
\end{document}